\documentclass[lettersize,journal]{IEEEtran}
\usepackage{amsmath,amsfonts}
\usepackage{algorithmic}
\usepackage{array}
\usepackage{subfig}
\usepackage{textcomp}
\usepackage{stfloats}
\usepackage{url}
\usepackage{verbatim}
\usepackage{graphicx}

\makeatletter

\newcommand{\Rmnum}[1]{\expandafter\@slowromancap\romannumeral #1@}
\makeatother

\usepackage{algorithm}
\usepackage{bm}
\usepackage{array}
\usepackage{makecell}
\usepackage{upgreek}
\usepackage{amssymb}

\def\BibTeX{{\rm B\kern-.05em{\sc i\kern-.025em b}\kern-.08em
    T\kern-.1667em\lower.7ex\hbox{E}\kern-.125emX}}
\usepackage{balance}

\begin{document}
\title{Joint Range-Velocity-Azimuth Estimation for OFDM-Based Integrated Sensing and Communication}
\author{Zelin Hu, Qibin Ye, Yixuan Huang, Su Hu, \textit{Member, IEEE}, and Gang Yang, \textit{Member, IEEE}
\thanks{This work was supported in part by the National Natural Science Foundation of China under Grants 61971092 and 62071093, in part by the Fundamental Research Funds for the Central Universities under Grant ZYGX2020ZB045 and Grant ZYGX2019J123, in part by the China Postdoctoral Science Foundation under Grant 2023M730508. \\ 
\indent The authors are with National Key Laboratory of Wireless Communications, University of Electronic Science and Technology of China, Chengdu 611731, China (e-mail: 202211220626@std.uestc.edu.cn; 202111220623@std.uestc.edu.cn; huangyx@uestc.edu.cn;husu@uestc.edu.cn; yanggang@uestc.edu.cn).}}

\markboth{}%
{How to Use the IEEEtran \LaTeX \ Templates}

\maketitle

\begin{abstract}
Orthogonal frequency division multiplexing (OFDM)-based integrated sensing and communication (ISAC) is promising for future sixth-generation mobile communication systems. Existing works focus on the joint estimation of the targets' range and velocity for OFDM-based ISAC systems. In contrast, this paper studies the three-dimensional joint estimation (3DJE) of range, velocity, and azimuth for OFDM-based ISAC systems with multiple receive antennas. First, we establish the signal model and derive the Cramér–Rao bounds (CRBs) on the 3DJE. Furthermore, an auto-paired super-resolution 3DJE algorithm is proposed by exploiting the reconstructed observation sub-signal's translational invariance property in the time, frequency, and space domains. Finally, with the 5G New Radio parameter setup, simulation results show that the proposed algorithm achieves better estimation performance and its root mean square error is closer to the root of CRBs than existing methods.
\end{abstract}

\section{Introduction}
\IEEEPARstart{I}{n} recent years, integrated sensing and communication (ISAC) has attracted fast-growing attention from both academia and industry \cite{ref1}-\cite{ref6}, as it has great application potential in various fields, such as smart home \cite{ref7}, internet of vehicles (IoV) \cite{ref8}, unmanned aerial vehicle \cite{ref9}, and action recognition \cite{ref10}. The ISAC systems integrate both sensing and communication functions into the same hardware platform and achieve the cooperation of radar sensing and communication, leading to remarkable advantages in terms of size, weight, cost, and power reduction, together with spectral efficiency improvement. Nowadays, ISAC has been considered as one of the main research directions of the sixth-generation (6G) mobile communication system \cite{ref11}.

Signal waveform design is an important aspect for ISAC systems \cite{ref12}. The current ISAC systems are mostly based on the existing communication, radar or emerging waveforms \cite{ref13}, such as frequency modulated continuous wave (FMCW) \cite{ref14}, orthogonal frequency division multiplexing (OFDM) \cite{ref15}, and orthogonal time frequency space (OTFS) \cite{ref16}. Waveform design can be divided into two main categories. The first category is the multiplexing waveform, which means that the communication and radar systems use different waveforms through time/frequency/space/code division multiplexing technologies \cite{ref17}-\cite{ref19}. The second category is the identity waveform, that is, the communication and radar systems share the same waveform. Since identity waveform achieves higher resource efficiency than multiplexing waveform \cite{ref20}, the identity waveform is considered in this paper.

Specifically, OFDM is a widely used communication waveform, which not only combats inter-symbol interference effectively but also has high spectral efficiency \cite{ref21}. The typical wireless systems that employ OFDM include wireless local area network \cite{ref22}, digital audio broadcasting \cite{ref23}, digital video broadcasting \cite{ref24}, and 5G New Radio (NR) \cite{ref25}. Meanwhile, since OFDM waveform has high range resolution and no range-Doppler coupling \cite{ref26}, it is also popular in radar applications, such as synthetic aperture radar \cite{ref27}, passive radar \cite{ref28}, and multiple-input multiple-output (MIMO) radar \cite{ref29}. Due to the above advantages, this paper considers the ISAC systems with OFDM waveforms. 

For OFDM-based ISAC systems, it is desirable to accurately estimate the target parameters (i.e., range, velocity, and azimuth). Existing works mainly discuss the joint range-and-velocity estimation for OFDM-based ISAC systems with single transmit and receive antenna. In \cite{ref12},  the most widely used two-dimensional discrete Fourier transform (2D-DFT) technique was proposed, which carried out discrete Fourier transform (DFT) operation on the Doppler axis and inverse discrete Fourier transform (IDFT) operation on the delay axis. The advantages of the 2D-DFT technique are low computational complexity and easy hardware implementation. However, for the 2D-DFT technique, the range resolution is limited by signal bandwidth, and the velocity resolution is limited by coherent processing interval (CPI). To improve range and velocity resolution, it requires to increase the signal bandwidth and CPI respectively, which is of high cost.

In fact, the received frequency-domain signal of the OFDM-based ISAC systems is analogous to the received space-domain signal for an array system with multiple receive antennas \cite{ref30}. Specifically, the received OFDM signal in the frequency domain is equivalent to the signal received by a virtual uniform rectangular array (URA) with one single snapshot, where each row and column of the virtual URA represent a subcarrier and a symbol, respectively. With the array signal model, some conventional super-resolution methods can be applied in joint estimation of range and velocity, such as the 2D multiple signal classification (2D-MUSIC) \cite{ref31}, 2D estimating signal parameter via rotational invariance techniques (2D-ESPRIT) and 2D-Capon \cite{ref32}. Moreover, since targets are sparse in the delay-Doppler domain, the compressive sensing theory was utilized to achieve joint estimation of range and velocity \cite{ref33}.

In this paper, we study the three-dimensional joint estimation (3DJE) of range, velocity and azimuth for OFDM-based ISAC systems with multiple receive antennas. Obviously, the above 2D estimation methods realize joint estimation of range and velocity; the classical direction of arrival (DOA) estimation methods, including DOA-Capon \cite{ref34}, DOA-MUSIC \cite{ref35} and DOA-ESPRIT \cite{ref36}, achieve estimation of azimuth. However, these methods are not applicable for 3DJE, because they cannot effectively pair the estimated azimuth with range and velocity in the case of multiple targets. Although the 3D-DFT method \cite{ref37} can be applied to 3DJE, it can not realize super-resolution estimation of parameters \cite{ref38}. To our best knowledge, the super-resolution 3DJE  problem for OFDM-based ISAC systems is seldomly studied in the literature. The main contributions of this paper are summarized as follows:

$\bullet$ We first establish the signal model for the OFDM-based ISAC systems with multiple receive antennas.  Specifically, the targets' ranges, velocities, and azimuths that cause different propagation delays, Doppler shifts, and wave paths, respectively, lead to phase rotations at the received echo signals. In addition, the Cramér–Rao bounds (CRBs) on the 3DJE are derived for such OFDM-based ISAC systems, which provide theoretical lower bounds for any unbiased 3DJE.

$\bullet$ We propose a two-step algorithm for 3DJE of range, velocity and azimuth, which not only achieves super resolution but also realizes automatic paring of multiple targets' ranges, velocities and azimuths. Specifically, the first step is the smoothing operation which removes the coherence of echo signals reflected by different targets and obtains the reconstructed observation sub-signal. The next step is to obtain the auto-paired parameter estimation by exploiting the reconstructed observation sub-signal's translational invariance property in the time, frequency and space domains. The fast subspace decomposition (FSD) of the reconstructed observation sub-signal's covariance matrix is conducted to obtain the signal subspace, leading to obvious complexity reduction. Besides, the computational complexity of the proposed 3DJE algorithm is analyzed. 

$\bullet$ Extensive numerical results are provided to verify the effectiveness of the proposed algorithm. The parameters of 5G NR standard are adopted. Simulation results show that the proposed 3DJE algorithm can significantly reduce the root-mean-square-error (RMSE) compared to classical algorithms, especially in the low signal-to-noise-ratio (SNR) region, and approaches the root of CRBs (RCRBs). Specifically,  for $\Delta f=120$ kHz, at a range RMSE level of 10$^{-1}$, the proposed algorithm achieves a SNR gain of 9.6 dB compared to the 2D-MUSIC method, and suffers from only 3.6 dB SNR degradation compared to the RCRB of range; at a velocity RMSE level of 10$^{-2}$, the proposed algorithm achieves an SNR gain of 4.2 dB compared to the 2D-MUSIC method, and suffers from only 2.8 dB SNR degradation compared to the RCRB of velocity; at an azimuth RMSE level of 10$^{-1}$, the proposed algorithm achieves a SNR gain of 2.9 dB compared to the DOA-MUSIC method, and suffers from only 1.6 dB SNR degradation compared to the RCRB of azimuth.

%the proposed algorithms suffers from slight RMSE increase from 0.04 to 0.05 for range estimation, from 0.02 to 0.03 for velocity estimation, from 0.065 to 0.075 for azimuth estimation, respectively, compared to the CRB. 
%the proposed algorithms reduces the RMSE by 100\%, 150\% and 65\%, compared to the the classic 2D-MUSIC and MUSIC algorithms, respectively. Moreover

The remaining parts of this paper are organized as follows. In Section \Rmnum{2}, the OFDM-based ISAC system model is presented. In Section \Rmnum{3}, the CRBs on the joint estimation of range, velocity and azimuth for OFDM-based ISAC systems are derived. In Section \Rmnum{4}, the proposed auto-paired super-resolution 3DJE algorithm is described. In Section \Rmnum{5}, simulation results are given. Section \Rmnum{6} concludes this paper.

Notations:${(  \cdot  )^T}$, ${(  \cdot  )^H}$ and  ${( \cdot )^\dag }$ stand for transpose, Hermitian transpose and pseudo inverse, respectively. $\otimes$ denotes Kronecker product, and $\odot$ denotes Hardamard product. $\rm{vec}\left(  \cdot  \right)$ represents matrix vectorization operation, ${\rm{rank}}\left(  \cdot  \right)$ represents matrix rank, $\det \left(  \cdot  \right)$ denotes the determinant, and ${\rm{angle}}\left(  \cdot  \right)$  represents the angle of complex. ${\mathop{\rm Re}\nolimits} \{  \cdot \}$ and ${\mathop{\rm Im}\nolimits} \{  \cdot \}$ stand for taking the real and imaginary part, respectively. $\mathbb{E}\left\{  \cdot  \right\}$ denotes the expectation, ${[ \cdot ]_{m,n}}$ denotes the $(m,n)$-th element of a matrix, and ${\left\|  \cdot  \right\|_2}$ denotes the Euclidean norm. ${\rm{diag(}}\mathbf x{\rm{)}}$ represents diagonal matrix with the elements of vector $\mathbf{x}$ on its diagonal, and ${\rm{Diag(}}\mathbf X{\rm{)}}$ represents the diagonal matrix formed by the diagonal elements of the square matrix $\mathbf{X}$. $ \mathbb{C}{^{M \times N}}$ is the set of $M \times N$ complex matrices. ${{\mathbf{I}}}$ is the identity matrix, and ${{\mathbf{I}}_i}$ denotes the \textit{i}-order identity matrix. ${{\bf{0}}_{i \times j}}$ represents the zero matrix with \textit{i} rows and \textit{j}  columns, and ${\hat{\mathbf{X}}}$ represents the estimation of $\mathbf{X}$.  ${\cal N}(\mu ,{\sigma ^2})$ and ${\cal C}{\cal N}(\mu ,{\sigma ^2})$ denote the Gaussian and the complex Gaussian distribution with mean $\mu$ and variance $\sigma^2$, respectively.

\section{System Model}
This section first describes the system model of OFDM-based ISAC briefly, and then presents the signal model for such systems.
\subsection{System Description}
\begin{figure}[!ht]
\centering
\includegraphics[width=3.2in]{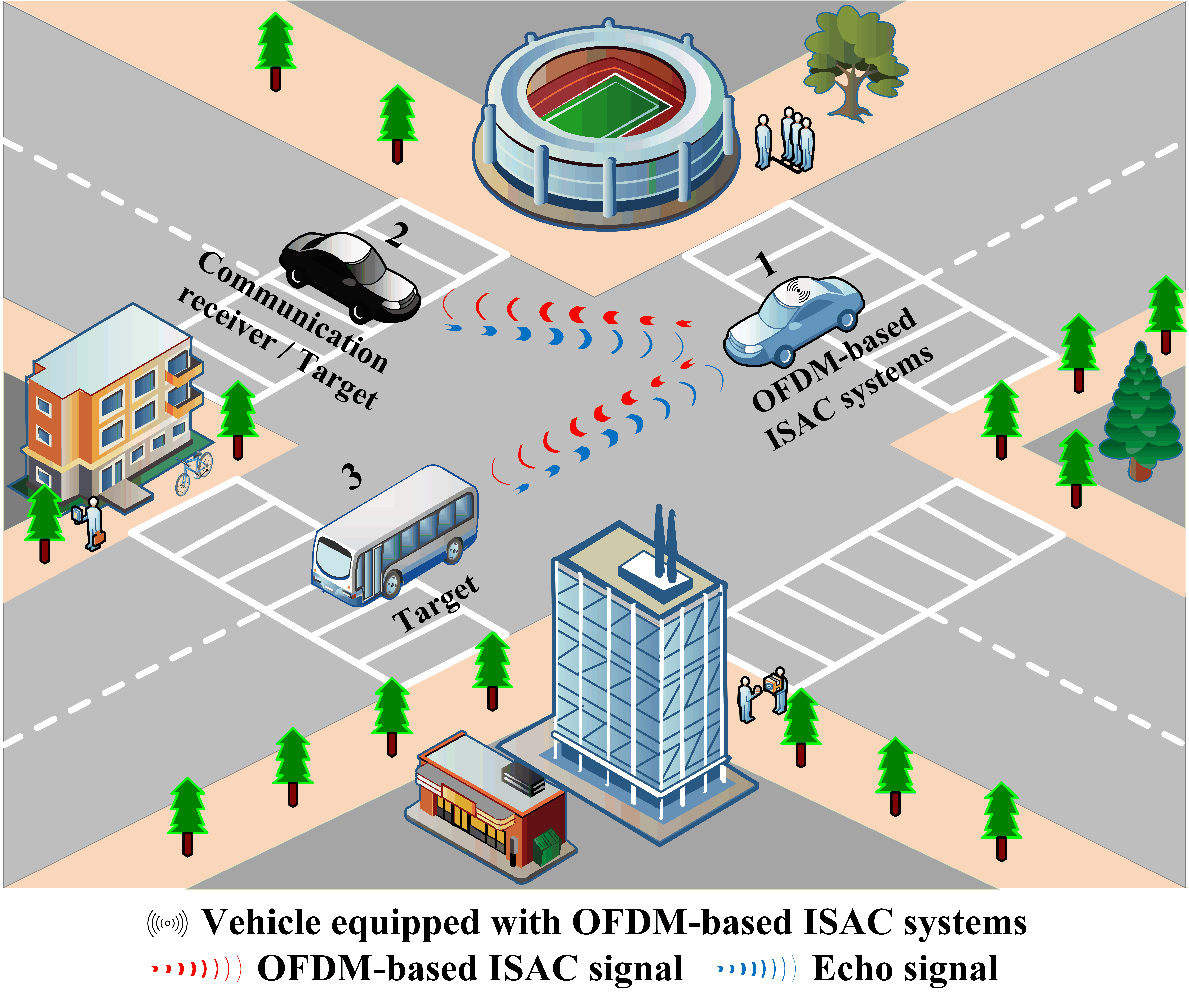}
\caption{IoV scenario for OFDM-based ISAC systems.}
\label{fig1}
\end{figure}

Fig. 1 illustrates a typical application scenario for OFDM-based ISAC systems in IoV. Vehicle 1 is equipped with an OFDM-based ISAC system, whereas vehicles 2 and 3 lack the ISAC function. Vehicle 1 transmits OFDM-based ISAC signal with a single antenna and receives echo signals reflected by the targets (i.e., vehicle 2 and vehicle 3) with multiple antennas. Suppose the transmit and receive antennas of vehicle 1 are spatially well-separated, and the self-leakage is ignored in this paper. The range, velocity and azimuth of vehicles 2 and 3 are estimated by processing the echo signals received at vehicle 1. Besides, vehicle 2 is also the communication receiver that can decode data symbols transmitted by vehicle 1. 
\subsection{Signal Model}
In the OFDM-based ISAC systems, the baseband OFDM signal with $ {N}$ $(N \geqslant 1)$ subcarriers within $\it{M}$ $(M \geqslant 1)$ OFDM symbol periods can be described as 
\begin{equation}
\label{deqn_ex1}
\begin{array}{ll}
x\!\left( t \right) &\!=\! \sum\limits_{m = 0}^{M \!-\! 1}\!x_m(t) \\
&\!=\!\sum\limits_{m = 0}^{M \!-\! 1} {\sum\limits_{n = 0}^{N \!-\! 1} {{s_m}\left( n \right)} {e^{j2\pi n\Delta f( {t\! -\! m\bar T} )}}\xi \!\left( {t \!-\! m\bar T} \right)} {\rm{,  }}
\end{array}
\end{equation}
where $x_m(t)$ is the baseband signal of the $m$-th OFDM symbol, ${s_m}\left( n \right)$ is the data symbol at the $n$-th subcarrier of the $m$-th OFDM symbol, $\Delta f$ is the subcarrier spacing, $\bar T = T + {T_{{\rm{cp}}}}$ is the duration of the completed OFDM symbol, $T={1 \mathord{\left/{\vphantom {1 {\Delta f}}} \right.\kern-\nulldelimiterspace} {\Delta f}}$ is the duration of the elementary OFDM symbol, ${T_{{\rm{cp}}}}$ is the duration of the cyclic prefix (CP), and the function $\xi \left( t \right)$ can be expressed as
\begin{equation}
\label{deqn_ex2}
\xi\left( t \right) = 
\begin{cases}
1,{\rm{}}t \in \left[{-T_{\rm{cp}},\bar T} \right],\\ 
0,{\rm{otherwise}}{\rm{.}} 
\end{cases}
\end{equation}

The baseband signal transmitted by a single antenna after upconversion can be written as 
\begin{equation}
\label{deqn_ex3}
\tilde x\left( t \right) = x\left( t \right){e^{j2\pi {f_{\rm{c}}}t}},
\end{equation}
where ${f_c}$ denotes the carrier frequency.%, which is much larger than the signal bandwidth.

We assume that the transmitted OFDM-based ISAC signal is reflected by $K$ $(K \geqslant 1)$ targets, whose range, velocity and azimuth are denoted as ${R_i},{v_i}$ and ${\theta _i}$, respectively, for $i\!=\! 0,1, \ldots ,K\!-\!1$. Accordingly, the delay of the $i$-th target is ${\tau _i} = {{2{R_i}} \mathord{\left/{\vphantom {{2{R_i}} c}} \right.\kern-\nulldelimiterspace} c}$, and the Doppler shift is ${f_{d,i}} = {{2{f_c}{v_i}} \mathord{\left/{\vphantom {{2{f_c}{v_i}} c}} \right.\kern-\nulldelimiterspace} c}$, where $c$ represents the speed of light. It should be noted that the maximum delay is less than the CP duration, i.e., $\mathop {\max }\limits_i \{ {\tau _i}\}  \le {T_{{\rm{cp}}}}$, and the maximum detectable range is 
\begin{equation}
\label{deqn_ex4}
{R_{\max }} = \frac{{c{T_{{\rm{cp}}}}}}{2}.
\end{equation}

The number of targets $K$ is assumed to be known, which can be obtained by using classic methods like the Akaike information criterion \cite{ref39} or minimum description length method \cite{ref40}.

\begin{figure}[!t]
\centering
\includegraphics[width=3in]{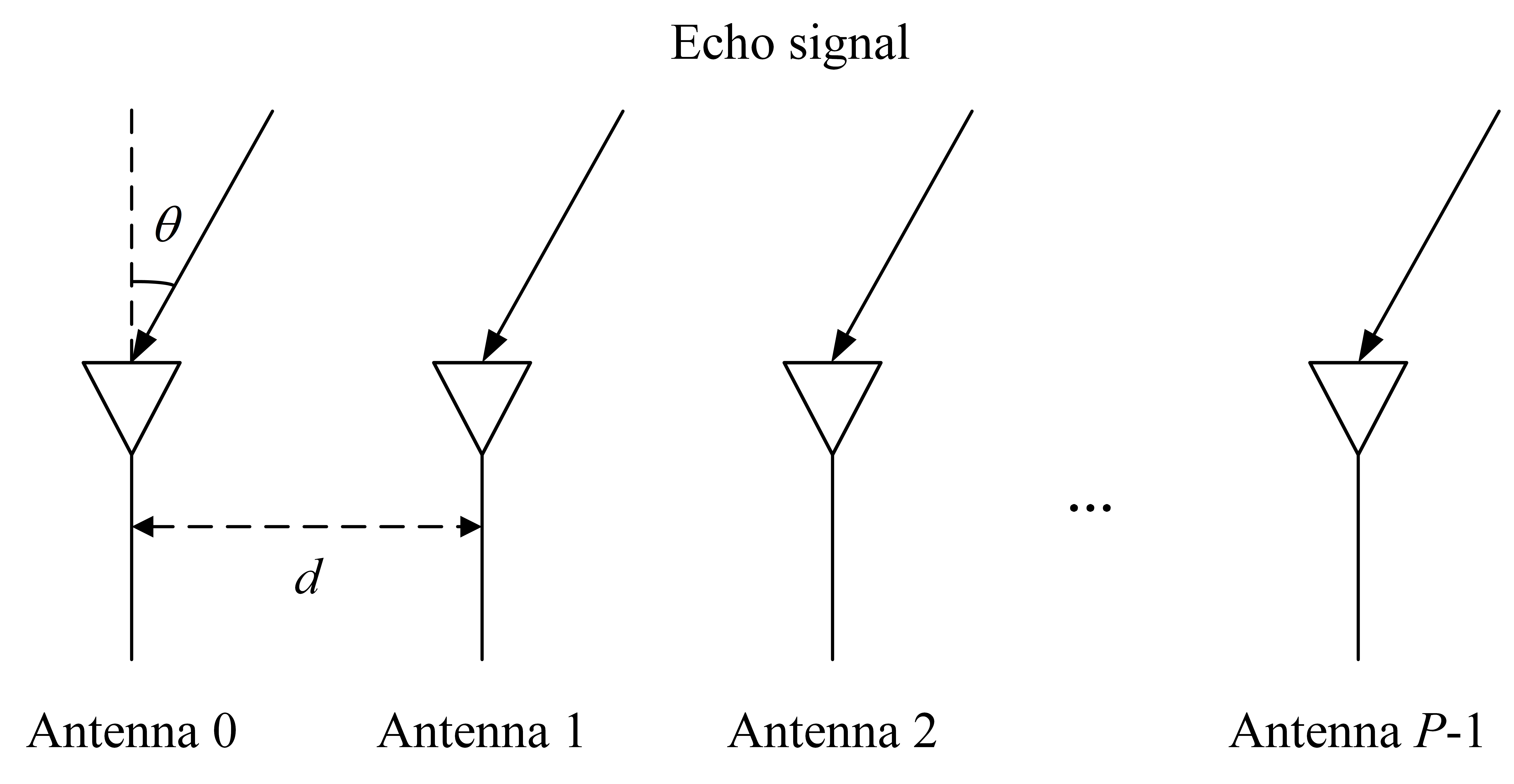}
\caption{Uniform linear array with inter-antenna spacing $d$.}
\label{fig2}
\end{figure}

The echo signals are received by a uniform linear array with inter-antenna spacing $d$, as shown in Fig. \ref{fig2}. Taking antenna 0 as the reference, the echo signal received by the $p$-th antenna ${\tilde y_p}(t)$, for $p = 0,1, \ldots ,P\!-\!1$, can be represented as 
\begin{equation}
\label{deqn_ex5}
{\tilde y_p}\left( t \right)\!=\sum\limits_{i = 0}^{K\!-\!1} {{\alpha _i}} \tilde x\left( {t - {\tau _i}} \right){e^{jp\frac{{2\pi d}}{\lambda }\!\sin \theta_i }}{e^{ j2\pi {f_{d,i}}t}} + {\tilde w_p}\left( t \right),
\end{equation}
where ${\alpha _i}$ denotes the amplitude of echo signal reflected by the $i$-th target, $\lambda = {c \mathord{\left/{\vphantom {c {{f_c}}}} \right.\kern-\nulldelimiterspace} {{f_c}}}$ is the wavelength, and ${\tilde w_p}\left( t \right)$ is the noise that is assumed to be additive white Gaussian noise with power ${\sigma ^2}$, i.e., ${\tilde w_p}\left( t \right)\sim{\cal C N}(0, {\sigma ^2})$.

After down-conversion, the signal in (\ref{deqn_ex5}) is converted to 
\begin{equation}
\label{deqn_ex6}
\begin{array}{lll}
{y_p}\!\left( t \right)  &\!\!\!\!\!\!= \!\!\sum\limits_{i = 0}^{K\!-\!1}\!\! {{\alpha _i}} x\left( {t \!-\! {\tau _i}} \right){e^{jp\frac{{2\pi d}}{\lambda }\!\sin \theta_i }}{e^{j2\pi {f_{d,i}}t}}{e^{ \!-\! j2\pi {f_c}{\tau _i}}}\!+\! {w'_p}\left( t \right) \\
 &\!\!\!\!\!\!=\!\! \sum\limits_{i = 0}^{K\!-\!1} \!\!{{\alpha_i'}} x\left( {t \!-\! {\tau _i}} \right){e^{jp\frac{{2\pi d}}{\lambda }\!\sin \theta_i }}{e^{ j2\pi {f_{d,i}}t}} + {w'_p}\left( t \right)\\
 &\!\!\!\!\!\!\approx \!\!\sum\limits_{i = 0}^{K\!-\!1}\sum\limits_{m = 0}^{M\!-\!1}\!\! {{\alpha_i'}} x_m\left( {t \!-\! {\tau _i}} \right)\!{e^{jp\frac{{2\pi d}}{\lambda }\!\sin \theta_i }}{e^{ j2\pi m{f_{d,i}}\bar T}} \!\!+\! {w'_p}\left( t \right),
\end{array}
\end{equation}
where  ${w'_p}\left( t \right) = {\tilde w_p}\left( t \right){e^{ - j2\pi {f_c}t}}$, ${w'_p}\left( t \right)\sim{\cal C N}(0 ,{\sigma ^2})$, and ${\alpha_i'} = {\alpha _i}{e^{ - j2\pi {f_c}{\tau _i}}}$. Since ${f_{d,i}}\bar T \ll 1$, the phase rotation within one OFDM symbol can be approximated as a constant, and the approximation in (\ref{deqn_ex6}) is valid \cite{ref30}, \cite{ref32}, \cite{ref33}.

Denote the sampling period ${T_s} = {1 \mathord{\left/{\vphantom {1 {N\Delta f}}} \right.\kern-\nulldelimiterspace} {(N\Delta f)}}$. The $N$  discrete samples can be obtained from each elementary OFDM symbol. Specifically, after discrete sampling and CP removal, the $k$-th discrete sample received at the $p$-th antenna in $m$-th OFDM symbol period can be written as
\begin{equation}
\label{deqn_ex7}
\begin{array}{l}
{y_{p,m}}\left[ k \right] \!=\! \sum\limits_{i = 0}^{K\!-\!1} {\sum\limits_{n = 0}^{N \!-\! 1} {{\alpha_i'}{s_m}\left( n \right){e^{jp\frac{{2\pi d}}{\lambda }\!\sin \theta_i }}{e^{j2\pi n\Delta f\left( {k{T_{\rm{s}}} - {\tau _i}} \right)}}} } \vspace{0.1cm} \\
\ \ \ \ \  \cdot \ {e^{j2\pi m{f_{d,i}}\bar T}} + {w'_{p,m}}\left( {k} \right),{\rm{  for }}\ k = 0,1, \ldots ,N\!-\!1.
\end{array}
\end{equation}
where ${w'_{p,m}}({k})\!=\!{w'_{p}}( {kT_s}+m\bar T )$, and ${w'_{p,m}}({k})\sim{\cal C N}(0 ,{\sigma ^2})$.
 
\indent The received symbol ${r_{p,m}}\left( n \right)$  is recovered from ${y_{p,m}}\left[ k \right]$ by a DFT operation and represented as
\begin{equation}
\label{deqn_ex8}
\begin{array}{l}
{r_{p,m}}\left( n \right) = \sum\limits_{k = 0}^{N \!-\! 1} {{y_{p,m}}[ k ]{e^{{{ - j2\pi nk} \mathord{\left/
 {\vphantom {{ - j2\pi nk} N}} \right.
 \kern-\nulldelimiterspace} N}}}} \\
 = \sum\limits_{i = 0}^{K\!-\!1} {{\beta _i}{s_m}\left( n \right){e^{jp\frac{{2\pi d}}{\lambda }\!\sin \theta_i }}{e^{j2\pi m{f_{{{d}},i}}\bar T}}{e^{ - j2\pi n\Delta f{\tau _i}}} \!+\! {w_{p,m}}\!\left( n \right)} \\
 = {s_m}\left( n \right)\sum\limits_{i = 0}^{K\!-\!1} {{\beta _i}{e^{jp\frac{{2\pi d}}{\lambda }\!\sin \theta_i }}{e^{j2\pi m{f_{{{d}},i}}\bar T}}{e^{ - j2\pi n\Delta f{\tau _i}}}\! +\! {w_{p,m}}\!\left ( n \right)} ,
\end{array}
\end{equation}
where \vspace{0.05cm} ${\beta _i} = N \alpha_i'$, and ${w_{p,m}}\left( n \right) \!=\! \sum\nolimits_{k = 0}^{N \!-\! 1} {{w'_{p,m}}({k}){e^{{{ - j2\pi nk} \mathord{\left/{\vphantom {{ - j2\pi nk} N}} \right.\kern-\nulldelimiterspace} N}}}} $ denotes the noise in the frequency domain.

For notational simplicity, we define the range, velocity and azimuth induced phases as ${\varphi _i} = 2\pi \Delta f{\tau_i}$, ${\phi _i} = 2\pi {f_{{{d,i}}}}\bar T$ and ${\psi_i} = {{2\pi d\sin {\theta _i}} \mathord{\left/
{\vphantom {{2\pi d\sin {\theta _i}} \lambda }} \right.
\kern-\nulldelimiterspace} \lambda }$, respectively, then ${r_{p,m}}\left( n \right)$ can be rewritten as 
\begin{equation}
\label{deqn_ex9}
{r_{p,m}}\left( n \right) \!= \!{s_m}\left( n \right)\!\sum\limits_{i = 0}^{K \!-\! 1} {{\beta _i}{e^{jp{\psi _i}}}{e^{jm{\phi _i}}}{e^{ - jn{\varphi _i}}} \!+\! {w_{p,m}}\left( n \right)}.
\end{equation}

For convenience, the received data symbol ${r_{p,m}}\left( n \right)$ is formed into matrix $\mathbf{D}_p \in \mathbb{C}^{M \times N}$, whose element in $m$-th row and $n$-th column is ${r_{p,m}}\left( n \right)$. The steering vectors are defined as ${{{\mathbf{a}}_D}( {\phi,m}) = {{[ {1,{e^{j\phi }}, \ldots ,{e^{j(m - 1)\phi }}} ]}^T}}$ and ${{{\mathbf{a}}_R}( {\varphi,n}) = {{[ {1,{e^{ - j\varphi }}, \ldots ,{e^{-j(n \!-\! 1)\varphi }}} ]}^T}}$. Thus, using the definitions ${{{\mathbf{A}}_D} = {{[ {{{\mathbf{a}}_D}( {{\phi _0},M} ),{{\mathbf{a}}_D}( {{\phi_1},M} ), \ldots ,{{\mathbf{a}}_D}( {{\phi_{K\!-\!1}},M} )} ]}}}$ and ${{{\mathbf{A}}_R} = {{[ {{{\mathbf{a}}_R}( {{\varphi_0},N}),{{\mathbf{a}}_R}( {{\varphi_1},N}) \ldots ,{{\mathbf{a}}_R}( {{\varphi_{K\!-\!1}},N} )}]}}}$, the received data matirx $\mathbf{D}_p$ is expressed as 
\begin{equation}
\label{deqn_ex10}
{\mathbf D_p} = {\mathbf D_t} \odot ({\mathbf A_D}{\rm diag}(\bm \upgamma_p )\mathbf A_R^T) +\mathbf  W_p,
\end{equation}
where ${{\bm{\upgamma }}_p} = {[ {{\beta _0}{e^{jp{\psi _0}}},{\beta _1}{e^{jp{\psi_1}}}, \ldots ,{\beta _{K\!-\!1}}{e^{jp{\psi _{K-1}}}}} ]^T}$,  $\mathbf D_t\!\in\! \mathbb{C}^{M \!\times\! N}$ and $\mathbf W_p\!\in \! \mathbb{C}^{M \!\times\! N}$ are matrices whose elements in $m$-th row and $n$-th column are $s_m(n)$ and ${w_{p,m}}\left( n \right)$, respectively.

In (\ref{deqn_ex10}), ${\bm{\upgamma }}_p$, ${\mathbf A_D}$ and ${\mathbf A_R}$ contain the targets' parameters to be estimated. In the receiver, the transmitted data symbol matrix ${\mathbf D_t}$ will interfere with parameter estimation. Since the originally transmitted data symbol ${\mathbf D_t}$ is known in advance by the ISAC systems itself, an element-wise division can be implemented to eliminate the influence of data symbols, i.e., $[\mathbf Z_p]_{m,n}=[{\mathbf D_p}]_{m,n}/[{\mathbf D_t}]_{m,n}$. Then, we have 

\begin{equation}
\label{deqn_ex11}
{{\mathbf{Z}}_p} = {{\mathbf{A}}_D}{\rm{diag}}\left( {{{\bm{\upgamma }}_p}} \right){\mathbf{A}}_R^T + {{\mathbf{V}}_p},
\end{equation}
where ${\mathbf{V}}_p$ is obtained by the element-wise division of ${\mathbf{W}}_p$ over ${\mathbf D_t}$. The element in $m$-th row and $n$-th column of ${\mathbf{V}}_p$ is ${v_{p,m}}({n})=[\mathbf V_p]_{m,n}=[{\mathbf W_p}]_{m,n}/[{\mathbf D_t}]_{m,n}$. \vspace{0.1cm}

\textit{Lemma 1}:  For the case of phase shift keying (PSK) symbol $s_m(n)$ and the original noise ${w'_{p,m}}( {k} )\sim {\cal C}{\cal N}(0 ,{\sigma ^2})$, the noise $v_{p,m}({n})$ \vspace{0.05cm} still follows a complex Gaussian distribution ${\cal C}{\cal N}(0 ,{\sigma_v^2})$, where $\sigma _v^2 = N{\sigma ^2}$. \vspace{0.1cm}

\textit{Proof}: Please refer to Appendix for details.\vspace{0.1cm}

Moreover, the vector ${{\mathbf{b}}_p} \buildrel \Delta \over = {\rm{vec}}\left( {{{\mathbf{Z}}_p}} \right)$  can be expressed as 
\begin{equation}
\label{deqn_ex12}
{{\mathbf{b}}_p} = \sum\limits_{i = 0}^{K\!-\!1} {{\beta _i}{e^{jp{\psi _i}}}{{\mathbf{a}}_R}\left( {{\varphi _i},N} \right) \otimes {{\mathbf{a}}_D}\left( {{\phi _i},M} \right)}  + {\rm{vec}}\left( {{{\mathbf{V}}_p}} \right).
\end{equation}

To use all the data symbols received by $P$ antennas, the vector ${\mathbf{b}} = {[ {{\mathbf{b}}_0;{\mathbf{b}}_1; \ldots ;{\mathbf{b}}_{P \!-\! 1}}]}\!\in\!\mathbb{C}^{M\!N\!P\times 1}$ can be written as
\begin{equation}
\label{deqn_ex13}
{\mathbf{b}} = \sum\limits_{i = 0}^{K\!-\!1} {\beta _i}{{\mathbf{a}}_S}\left( {{\psi _i},P} \right)  \otimes {{\mathbf{a}}_R}\left( {{\varphi _i},N} \right) \otimes {{{\mathbf{a}}_D}\left( {{\phi _i},M} \right)} + {\mathbf{v}}, 
\end{equation}
where \vspace{0.05cm} ${\mathbf{v}} \!=\! [ {\rm{vec}} ({{\mathbf{V}}_0});  {\rm{vec}} ({{\mathbf{V}}_1}); \ldots; {\rm{vec}}({\mathbf{V}}_{P \!-\! 1})]$ and ${{\mathbf{a}}_S}( {\psi, p} ) = {[ {1,{e^{j\psi }}, \ldots ,{e^{j(p - 1)\psi }}} ]^T}$.

For simplicity, define vectors ${\bm {\upbeta}} ={[{\beta _0},{\beta _1}, \ldots ,{\beta _{K \!- \!1}}]^T}$ and $\mathbf a_i={{\mathbf{a}}_S}\left( {{\psi_i},P} \right)  \otimes {{\mathbf{a}}_R}\left( {{\varphi _i},N} \right) \otimes {{{\mathbf{a}}_D}\left( {{\phi _i},M} \right)}$. Then, ${\mathbf{b}}$ can be further expressed as
\begin{equation}
\label{deqn_ex14}
{\mathbf{b}} = \mathbf{A}\bm \upbeta  + \mathbf v,
\end{equation}
where $\mathbf A =[{\mathbf a_0},{\mathbf a_1}, \ldots, {\mathbf a_{K\!-\!1}}]$.

Thus, the sensing problem is to jointly estimate the unknown parameters $\mathbf{A}$ and $\bm{\upbeta}$ from the noisy observation ${\mathbf{b}}$. With these estimations, the range ${R_i}$, velocity ${v_i}$, and azimuth ${\theta_i}$ can be calculated directly.

\section{CRBs on Range Velocity and Azimuth Estimation}
In this section, the CRBs on the joint estimation of range velocity and azimuth are derived.

The parameters that need to be estimated in (\ref{deqn_ex14}) are 
\begin{equation}
\label{deqn_ex15}
\bm \Theta  = {\left[ {{\sigma_v^2},{\mathop{\rm Re}\nolimits} {{{\rm{\{ }}\bm \upbeta {\rm{\} }}}^T},{\mathop{\rm Im}\nolimits} {{{\rm{\{ }}\bm \upbeta {\rm{\} }}}^T},{\bm \upvarphi ^T},\bm \upphi^T,{\bm \uptheta ^T}} \right]^T},
\end{equation}
where \vspace{0.1cm}${\mathop{\rm Re}\nolimits} {\rm{\{ }}\bm \upbeta {\rm{\} }} = {\left[ {{\mathop{\rm Re}\nolimits} \{ {\beta _0}\}, \ldots ,{\mathop{\rm Re}\nolimits} \{ {\beta _{K \!-\! 1}}\} } \right]^T}$, ${\mathop{\rm Im}\nolimits} {\rm{\{ }}\bm \upbeta {\rm{\} }} = {\left[ {{\mathop{\rm Im}\nolimits} \{ {\beta _0}\},\ldots , {\mathop{\rm Im}\nolimits} \{ {\beta _{K\!-\!1}}\} } \right]^T}, \vspace{0.1cm} \bm \upvarphi  = {[{\varphi _0},{\varphi _1},\ldots ,{\varphi _{K\!-\!1}}]^T}, {\bm \upphi} = {[\phi_0,\phi_1\ldots ,\phi_{K\!-\!1}]^T}, \bm \uptheta  = {[{\theta_0},{\theta _1},\ldots ,{\theta_{K\!-\!1}}]^T}$.

\textit{Theorem 1}: With the signal model in (\ref{deqn_ex14}), the CRBs on the joint estimation of range, velocity and azimuth are given by 
\begin{equation}
\label{deqn_ex16}
{\rm{CRB}}(\bar {\bm \Theta}) = \frac{{{\sigma_v^2}}}{2}{\left[ {{\mathop{\rm Re}\nolimits} \{ ({\bar{\mathbf A}^H}\mathbf P_{\mathbf A}^ \bot \bar{\mathbf A})  \odot {\mathbf \Omega ^T\}}} \right]^{ \!-\! 1}},
\end{equation}
where 
\begin{equation}
\label{deqn_ex17}
\begin{array}{c}
 \bar {\bm \Theta}  = {[{\bm \upvarphi^T},\bm \upphi^T,{\bm \uptheta ^T}]^T},\vspace{0.15cm}\\
{\mathbf P}_{\mathbf A}^ \bot  = {\mathbf I} - {\mathbf A}{({{\mathbf A}^H}{\mathbf A})^{ \!-\! 1}}{\mathbf A}, \vspace{0.15cm}\\
{\mathbf{\Omega }} = \left[ {\begin{array}{*{20}{c}}
{{\bm{\upbeta }}{{\bm{\upbeta }}^H}}&{{\bm{\upbeta }}{{\bm{\upbeta }}^H}}&{{\bm{\upbeta }}{{\bm{\upbeta }}^H}}\\
{{\bm{\upbeta }}{{\bm{\upbeta }}^H}}&{{\bm{\upbeta }}{{\bm{\upbeta }}^H}}&{{\bm{\upbeta }}{{\bm{\upbeta }}^H}}\\
{{\bm{\upbeta }}{{\bm{\upbeta }}^H}}&{{\bm{\upbeta }}{{\bm{\upbeta }}^H}}&{{\bm{\upbeta }}{{\bm{\upbeta }}^H}}\\
\end{array}} \right],\vspace{0.1cm}\\ 
{\bar{\mathbf{A}}} =\bigg[\displaystyle \frac{{\partial {{\mathbf{a}}_0}}}{{\partial {\varphi_0}}},\frac{{\partial {{\mathbf{a}}_1}}}{{\partial {\varphi_1}}}, \ldots ,\frac{{\partial {{\mathbf{a}}_{K\!-\!1}}}}{{\partial {\varphi_{K\! - \!1}}}},\frac{{\partial {{\mathbf{a}}_0}}}{{\partial {\phi_0}}}, \frac{{\partial {{\mathbf{a}}_1}}}{{\partial {\phi_1}}}, \ldots, \frac{{\partial {{\mathbf{a}}_{K\!-\!1}}}}{\partial {\phi_{K\!-\!1}}}, \vspace{0.1cm}  \\ \ \ \ \ \ \
\displaystyle \frac{{\partial {{\mathbf{a}}_0}}}{{\partial {\theta_0}}},\frac{{\partial {{\mathbf{a}}_1}}}{{\partial {\theta_1}}}, \ldots , \frac{{\partial {{\mathbf{a}}_{K\!-\!1}}}}{{\partial {\theta_{K\!-\!1}}}} \bigg].
\end{array}
\end{equation}
\textit{Proof}: From (\ref{deqn_ex14}), the likelihood function is
\begin{equation}
\label{deqn_ex18}
L = \displaystyle \frac{1}{{{\pi ^{MNP}}{\sigma_v ^{2MNP}}}}\exp \left\{  - \frac{1}{{{\sigma_v^2}}}\left\| {\mathbf b - \mathbf A \bm \upbeta } \right\|_2^2 \right \}.
\end{equation}\par
Then, the logarithmic likelihood function is
\begin{equation}
\label{deqn_ex19}
\ln\! L =  - M\!N\!P\ln \pi  - M\!N\!P\ln {\sigma_v ^2} - \displaystyle \frac{1}{{{\sigma_v^2}}}\left\| {\mathbf b - \mathbf A \bm \upbeta } \right\|_2^2.
\end{equation}\par

Since ${\rm{CRB}}(\mathbf \Theta)= \mathbf{\bar F}^{\!-\!1}$, where $\mathbf {\bar F}$ is the Fisher information matrix, that can be written as 
\begin{equation}
\label{deqn_ex20}
\mathbf {\bar F} = {\rm{\mathbb{E}}} \left\{ \frac{{\partial \ln \!L}}{{\partial \mathbf \Theta }}\frac{{\partial \ln \!L}}{{\partial {\mathbf \Theta ^T}}} \right\}.
\end{equation}\par

The partial derivative over the parameters ${\sigma_v^2}$, ${\mathop{\rm Re}\nolimits} {{{\rm{\{ }}\bm \upbeta {\rm{\} }}}}$, ${\mathop{\rm Im}\nolimits}{{{\rm{\{ }}\bm \upbeta {\rm{\} }}}}$, and ${\bm \upvarphi }\bm$  are derived as follows:
\begin{equation}
\label{deqn_ex21}
\frac{{\partial \ln \!L}}{{\partial {\sigma_v^2}}} =  - \frac{{M\!N\!P}}{{{\sigma_v^2}}} + \frac{1}{{{\sigma_v^4}}}\left\| {\mathbf b - \mathbf A \bm \upbeta } \right\|_2^2,
\end{equation}
\begin{equation}
\label{deqn_ex22}
\begin{array}{ll}
\displaystyle \frac{{\partial \ln \!L}}{{\partial {\mathop{\rm Re}\nolimits} \{ \bm \upbeta \} }} & \!\!\!= {\left[\displaystyle \frac{{\partial \ln\! L}}{{\partial {\mathop{\rm Re}\nolimits} \{ {\beta _0}\} }}, \ldots,\displaystyle \frac{{\partial \ln \!L}}{{\partial {\mathop{\rm Re}\nolimits} \{ {\beta _{K\!-\!1}}\} }}\right]^T} \vspace{0.1cm}\\
&\!\!\!= \displaystyle \frac{1}{{{\sigma_v^2}}}[{\mathbf A^H}\mathbf v + {\mathbf A^T}{\mathbf v^ * }] \vspace{0.1cm} \\
&\!\!\!= \displaystyle \frac{2}{{{\sigma_v^2}}}{\mathop{\rm Re}\nolimits} \{ {\mathbf A^H}\mathbf v\},
\end{array}
\end{equation}
\begin{equation}
\label{deqn_ex23}
\begin{array}{ll}
\displaystyle \frac{{\partial \ln\! L}}{{\partial {\mathop{\rm Im}\nolimits} \{ \bm \upbeta \} }} &\!\!\!= {\left[\displaystyle \frac{{\partial \ln\! L}}{{\partial {\mathop{\rm Im}\nolimits} \{ {\beta _0}\} }},\ldots,\displaystyle \frac{{\partial \ln L}}{{\partial {\mathop{\rm Im}\nolimits} \{ {\beta _{K \!-\! 1}}\} }}\right]^T} \vspace{0.1cm} \\
&\!\!\!= \displaystyle \frac{1}{{{\sigma_v^2}}}[ - j{\mathbf A^H}\mathbf v + j{\mathbf A^T}{\mathbf v^ * }] \vspace{0.1cm} \\
&\!\!\!= \displaystyle \frac{2}{{{\sigma_v^2}}}{\mathop{\rm Im}\nolimits} \{ {\mathbf A^H}\mathbf v\},
\end{array}
\end{equation}
\begin{equation}
\label{deqn_ex24}
\displaystyle \frac{{\partial \!\ln \!L}}{{\partial \bm \upvarphi }}\!\!=\!\! {\left[\!\displaystyle \frac{{\partial \! \ln\! L}}{{\partial {\varphi_0}}},\! \displaystyle \frac{{\partial \! \ln\! L}}{{\partial {\varphi_1}}},\!\ldots \!,\displaystyle \frac{{\partial \! \ln \!L}}{{\partial {\varphi_{K\!-\!1}}}}\!\right]^T} \!\!\!\!\!=\!\! \displaystyle \frac{2}{{{\sigma_v^2}}}{\mathop{\rm Re}\nolimits} {\rm{\{ }}{\mathbf B^H}\!\bar{\mathbf A}_1^H \!\mathbf v{\rm{\} }}, \end{equation} 
where $\mathbf B = \rm{diag}(\bm \upbeta )$ and ${\bar {\mathbf{A}}_1} = \left[\displaystyle \frac{{\partial {{\mathbf{a}}_0}}}{{\partial {\varphi_0}}}, \frac{{\partial {{\mathbf{a}}_1}}}{{\partial {\varphi_1}}}, \ldots ,\frac{{\partial {{\mathbf{a}}_{K\!-\!1}}}}{{\partial {\varphi _{K\!-\!1}}}}\right]$.\vspace{0.05cm}

The partial derivative over the parameters $\bm \upphi$ and $\bm \uptheta$ are derived as follows:
\begin{equation}
\label{deqn_ex25}
\displaystyle \frac{{\partial \!\ln\! L}}{{\partial {\bm \upphi}}}\!\!= \!\!{\left[ \! \displaystyle \frac{{\partial \!\ln\! L}}{{\partial {\phi_0}}},\! \displaystyle \frac{{\partial \!\ln\! L}}{{\partial {\phi_1}}},\!\ldots ,\!\displaystyle \frac{{\partial \!\ln \!L}}{{\partial {\phi_{K\!-\!1}}}}\! \right]^T}
\!\!\!= \!\!\displaystyle\frac{2}{{{\sigma_v^2}}}{\mathop{\rm Re}\nolimits} {\rm{\{ }}{\mathbf B^H} \bar{\mathbf A}_2^H \mathbf v{\rm{\} }}, 
\end{equation}
\begin{equation}
\label{deqn_ex26}
\displaystyle \frac{{\partial \!\ln \!L}}{{\partial \bm \uptheta }} \!\!= {\left[\!\displaystyle \frac{{\partial \! \ln\! L}}{{\partial {\theta _0}}},\!\frac{{\partial \!\ln \! L}}{{\partial {\theta _1}}}, \!\ldots \!,\!\displaystyle \frac{{\partial \! \ln \!L}}{{\partial {\theta _{K\!-\!1}}}}\right]^T}\!\!\!=\!\!\displaystyle \frac{2}{{{\sigma_v^2}}}{\mathop{\rm Re}\nolimits} {\rm{\{ }}{\mathbf B^H}\bar{\mathbf A}_3^H \mathbf v{\rm{\} }},
\end{equation}
where $\bar{\mathbf A}_2 \!=\! \left[\!\displaystyle \frac{{\partial {\mathbf a_0}}}{{\partial {\phi_0}}}, \ldots ,\displaystyle \frac{{\partial {\mathbf a_{K\!-\!1}}}}{{\partial {\phi_{K\!-\!1}}}}\!\right]$ and  ${\bar {\mathbf A}_3} \!=\! \left[\!\displaystyle \frac{{\partial {\mathbf a_0}}}{{\partial {\theta _0}}}, \ldots ,\displaystyle \frac{{\partial {\mathbf a_{K\!-\!1}}}}{{\partial {\theta _{K\!-\!1}}}}\!\right]$.\vspace{0.1cm}

\newcounter{TempEqCnt} % 创建临时变量TempEqCnt
\setcounter{TempEqCnt}{\value{equation}} % 将当前公式序号 赋给TempEqCnt
\setcounter{equation}{34} % 当前公式序号变为x，x等于长公式应有的序号减1.
\begin{figure*}[!b]
 \hrulefill
    \begin{equation}
    \label{deqn_ex35}
{{\mathbf{\tilde b}}_{\tilde p+l,\tilde m,\tilde n}} = \rm{vec} \left(\left[ {\begin{array}{*{20}{c}}
\!{{z_{\tilde p+l,\tilde m}}(\tilde n)}&\!{{z_{\tilde p+l,\tilde m}}(\tilde n + 1)}& \!\!\cdots &{{z_{\tilde p+l,\tilde m}}(\tilde n + \tilde N - 1)}\\
\!{{z_{\tilde p+l,\tilde m + 1}}(\tilde n)}&\!{{z_{\tilde p+l,\tilde m + 1}}(\tilde n + 1)}& \!\!\cdots &{{z_{\tilde p+l,\tilde m + 1}}(\tilde n + \tilde N - 1)}\\
 \vdots & \vdots & \ddots & \vdots \\
\!{{z_{\tilde p+l,\tilde m + \tilde M - 1}}(\tilde n)}&\!{{z_{\tilde p+l,\tilde m + \tilde M - 1}}(\tilde n + 1)}& \!\!\cdots &{{z_{\tilde p+l,\tilde m + \tilde M - 1}}(\tilde n + \tilde N - 1)}
\end{array}} \right] \right)
    \end{equation}
\end{figure*}
\setcounter{equation}{\value{TempEqCnt}} 
\setcounter{equation}{26} % 当前公式序号变为y，y等于长公式的序号. 如果是有两个 % 或两个以上的长公式，那么y是指最后一个长公式的序号。

Since $ \bar {\bm \Theta}  = {[{\bm \upvarphi^T},\bm \upphi^T,{\bm \uptheta ^T}]^T}$, we have  
\begin{equation}
\label{deqn_ex27}
\begin{array}{l}
\displaystyle \frac{{\partial \ln \!L}}{{\partial \bar{\boldsymbol \Theta}}}\!=\! {\left[\displaystyle \frac{{\partial \ln\! L}}{{\partial {\bm \upvarphi ^T}}},\!\displaystyle \frac{{\partial \ln\! L}}{{\partial \bm \upphi^T}},\!\displaystyle \frac{{\partial \ln \! L}}{{\partial {\bm \uptheta ^T}}}\right]^T}\!\!\!=\!\displaystyle \frac{2}{{{\sigma_v^2}}}{\mathop{\rm Re}\nolimits} {\rm{\{ }}{{ \tilde{\mathbf B}}^H}\!{\mathbf {\bar A}^H}\!\mathbf v{\rm{\} }},
\end{array}
\end{equation}
where $\tilde{\mathbf B} = \left[ {\begin{array}{*{20}{c}}
{\mathbf B}&{\mathbf 0}&{\mathbf 0}\\
{\mathbf 0}&{\mathbf B}&{\mathbf 0}\\
{\mathbf 0}&{\mathbf 0}&{\mathbf B}
\end{array}} \right]$.\vspace{0.2cm} \par
From Theorem 4.1 in \cite{ref40}, we can obtain
\begin{equation}
\begin{array}{ll}
{\rm{CRB}}(\bar {\mathbf \Theta}) & \!\!\!= \displaystyle \frac{{{\sigma_v^2}}}{2}{\left[ {{\mathop{\rm Re}\nolimits} \{ \mathbf{\tilde B}^H( {\bar{\mathbf A}^H}\mathbf P_{\mathbf A}^ \bot \bar{\mathbf A}) \mathbf {\tilde B} \}} \right]^{\! - \!1}} \vspace{0.1cm}\\
& \!\!\!=\displaystyle \frac{{{\sigma_v^2}}}{2}{\Big[ {{\mathop{\rm Re}\nolimits} \{ ({\bar{\mathbf A}^H}\mathbf P_{\mathbf A}^ \bot \bar{\mathbf A})  \odot {\mathbf \Omega ^T\}}} \Big]^{ \!-\! 1}},\nonumber
\end{array}
\end{equation}
where ${\mathbf  P}_{\mathbf  A}^ \bot$, $\bar{\mathbf  A}$ and ${\mathbf {\Omega }}$ are defined in (\ref{deqn_ex17}). The proof is thus complete. $\hfill\blacksquare$ \vspace{0.1cm} \par
The CRBs of $\varphi_i$, ${\phi_{i}}$, and azimuth $\theta_i$ are obtained as
\begin{equation}
\label{deqn_ex28}
{\rm{CRB}}({\varphi _i}) = {\left[ {{\rm{CRB}}(\bar {\boldsymbol \Theta} )} \right]_{i,i}},
\end{equation}
\begin{equation}
\label{deqn_ex29}
{\rm{CRB}}(\phi_{i}) = {\left[ {{\rm{CRB}}(\bar {\boldsymbol \Theta})} \right]_{K + i,K + i}},
\end{equation}
\begin{equation}
\label{deqn_ex30}
{\rm{CRB}}({\theta_{i}}) = {\left[ {{\rm{CRB}}(\bar {\boldsymbol \Theta})} \right]_{2K + i,2K + i}},
\end{equation}
Furthermore, the CRBs on range $R_i$ and velocity $v_i$ can be written as
\begin{equation}
\label{deqn_ex31}
{\rm{CRB}}({R_i}) = \frac{{{c^2}}}{(4\pi\Delta f)^2}{\rm{CRB}}({\varphi _i}),
\end{equation}
\begin{equation}
\label{deqn_ex32}
{\rm{CRB}}({v_i}) = \frac{{{\lambda ^2}}}{(4\pi{ \bar T})^2}{\rm{CRB}}(\phi_{i}).
\end{equation}

\section{Auto-Paired Super-Resolution 3DJE}
This section presents the proposed 3DJE algorithm, including smoothing operation and auto-paired super-resolution estimation.
\subsection{Smoothing Operation}
The signal in (\ref{deqn_ex14}) is equivalent to the classic array signal model, whose steering vector is $\mathbf a_i={{\mathbf{a}}_S}\left( {{\psi _i},P} \right)  \otimes {{\mathbf{a}}_R}\left( {{\varphi _i},N} \right) \otimes {{{\mathbf{a}}_D}\left( {{\phi _i},M} \right)}$. Nevertheless, the equivalent array signal has only one snapshot, resulting in the coherence of echo signals reflected by different targets. As a consequence, the smoothing operation is necessary for the decoherence.

For simplicity, we redefine the steering vector 
\begin{equation}
\label{deqn_ex33}
{\tilde{\mathbf{a}}_i}(\!\tilde P,\!\tilde M,\!\tilde N\!) \buildrel \Delta \over = {{\mathbf{a}}_S}( {{\psi_i},\tilde P} )  \otimes {{\mathbf{a}}_R}( {{\varphi_i},\tilde N} ) \otimes {{{\mathbf{a}}_D}( {{\phi_i},\tilde M} )}\in\mathbb{C}^{N_{{{L}}}\times 1},
\end{equation}
where $\tilde N \!<\! N,\tilde M\! <\! M,\tilde P \!<\! P$ and ${N_{{{L}}}} \!= \!{\tilde M\!\tilde N\!\tilde P}$. Therefore, there are ${N_{{\rm{s}}}} \!= \! ({N \!-  \!\tilde N \! + \! 1} )({M \!-\! \tilde M \!+ \!1} )({P \!- \!\tilde P \!+ \!1} )$\! \!\!\! snapshots after smoothing operation.

Specifically, the original observation signal $\mathbf{b}\!\in\!\mathbb{C}^{M\!N\!P\times 1}$ is reconstructed into $N_{\rm{s}}$ observation sub-signals (snapshots), and the reconstructed observation sub-signals can be expressed as  
\begin{equation}
\label{deqn_ex34}
\begin{array}{ll}
{\mathbf g}_{\tilde p,\tilde m,\tilde n} \!\! \buildrel \Delta \over = \! {[ {\mathbf{\tilde b}_{\tilde p,\tilde m,\tilde n};\!\mathbf {\tilde b}_{\tilde p +\! 1,\tilde m,\tilde n}; \ldots ;\!\mathbf {\tilde b}_{\tilde p + \!\tilde P \!-\! 1,\tilde m,\tilde n}}]}\!\in\!\mathbb{C}^{N_{{{L}}}\!\times\! 1},
\end{array}
\end{equation}
where \vspace{0.1cm}$\tilde p \!= \! 0,1, \ldots ,P\!-\!\tilde P, \tilde m \!=\! 0,1, \ldots ,M \!-\! \tilde M,{\rm{ }}\tilde n \!=\! 0,1, \ldots ,N \!-\! \tilde N$, and ${{\mathbf{\tilde b}}_{\tilde p+l,\tilde m,\tilde n}}$ is given by (\ref{deqn_ex35}) shown at the bottom of this page with $l = 0,1, \ldots ,\tilde P\!-\!1$.

\setcounter{equation}{36} % 当前公式序号变为x，x等于长公式应有的序号减1.
\begin{figure*}[!ht]
    \begin{equation}
    \label{deqn_ex37}
{{\mathbf{\tilde v}}_{\tilde p+l,\tilde m,\tilde n}} = \rm{vec}\left(\left[ {\begin{array}{*{20}{c}}
\!{{v_{\tilde p+l,\tilde m}}(\tilde n)}&\!{{v_{\tilde p+l,\tilde m}}(\tilde n + 1)}& \!\!\cdots &{{v_{\tilde p+l,\tilde m}}(\tilde n + \tilde N - 1)}\\
\!{{v_{\tilde p+l,\tilde m + 1}}(\tilde n)}&\!{{v_{\tilde p+l,\tilde m + 1}}(\tilde n + 1)}& \!\!\cdots &{{v_{\tilde p+l,\tilde m + 1}}(\tilde n + \tilde N - 1)}\\
 \vdots & \vdots & \ddots & \vdots \\
\!{{v_{\tilde p+l,\tilde m + \tilde M - 1}}(\tilde n)}&\!{{v_{\tilde p+l,\tilde m + \tilde M - 1}}(\tilde n + 1)}& \!\!\cdots &{{v_{\tilde p+l,\tilde m + \tilde M - 1}}(\tilde n + \tilde N - 1)}
\end{array}} \right] \right)
    \end{equation}
     \hrulefill
\end{figure*}

Furthermore, $\mathbf g_{\tilde p,\tilde m,\tilde n}$ can be expressed as 
\setcounter{equation}{35} % 当前公式序号变为y，y等于长公式的序号. 如果是有两个 % 或两个以上的长公式，那么y是指最后一个长公式的序号。
\begin{equation}
\label{deqn_ex36}
{\mathbf g}_{\tilde p,\tilde m,\tilde n} ={\tilde{\mathbf A}}{\rm{diag(}}\bm \upbeta {\rm{)}}{\mathbf x_{\tilde p,\tilde m,\tilde n}} + {\mathbf{n}_{\tilde p,\tilde m,\tilde n}},
\end{equation}
where ${\tilde{\mathbf A}} = [ {\tilde{\mathbf {a}}_0}(\!\tilde P,\!\tilde M,\!\tilde N\!), {\tilde{\mathbf{a}}_1}(\!\tilde P,\!\tilde M,\!\tilde N\!), \ldots {\tilde{\mathbf{a}}_{K\!-\!1}}(\!\tilde P,\!\tilde M,\!\tilde N\!)]$ \vspace{0.1cm}, ${\mathbf x_{\tilde p,\tilde m,\tilde n}} = {[ {{e^{ j(\tilde m{\phi _0} - \tilde n{\varphi _0} + \tilde p{\psi _0})}}, \ldots ,{e^{j (\tilde m{\phi _{K\!-\!1}} - \tilde n{\varphi _{K\!-\!1}} + \tilde p{\psi _{K\!-\!1}})}}} ]^T}$, and ${\mathbf n_{\tilde p,\tilde m,\tilde n}}\! \!=\!\! {[ {{{\mathbf{\tilde v}}^T}_{\tilde p,\tilde m,\tilde n},{{\mathbf{\tilde v}}^T}_{\tilde p + 1,\tilde m,\tilde n}, \ldots ,{{\mathbf{\tilde v}}^T}_{\tilde p + \tilde P - 1,\tilde m,\tilde n}} ]^T}$ is obtained by smoothing operation on $\mathbf v$, and ${{\mathbf{\tilde v}}_{\tilde p+l,\tilde m,\tilde n}}$ is shown in (\ref{deqn_ex37}) with $l = 0,1, \ldots ,\tilde P\!-\!1$.
\subsection{Auto-Paired Super-Resolution Estimation}
After smoothing operation, the reconstructed observation sub-signal in (\ref{deqn_ex36}) is still equivalent to the classic array model but with $N_{\rm{s}}$ snapshots. Then, the auto-paired super-resolution 3DJE is realized in this subsection by utilizing the translational invariance property of the sub-signal model in the time, frequency, and space domains.
 
First, the covariance matrix of the reconstructed observation sub-signal ${\mathbf g}_{\tilde p,\tilde m,\tilde n}$ can be written as
\setcounter{equation}{37}
\begin{equation}
\label{deqn_ex38}
\begin{array}{ll}
\boldsymbol{\Psi}=\mathrm{\mathbb{E}}\left\{ {{\mathbf g_{\tilde p,\tilde m,\tilde n}}\mathbf g_{\tilde p,\tilde m,\tilde n}^H} \right\} = {\tilde{\mathbf A}}\boldsymbol{\Gamma} {\tilde{\mathbf A}}^H + {\sigma_v^2}{\mathbf I_{{N_L}}},
\end{array}
\end{equation}
where $\boldsymbol \Gamma  = {\rm{diag(}}\bm \upbeta {\rm{)}}\mathrm{\mathbb{E}}\left\{ {{\mathbf x_{\tilde p,\tilde m,\tilde n}}\mathbf x_{\tilde p,\tilde m,\tilde n}^H} \right\}{\left( {{\rm{diag(}}\bm \upbeta {\rm{)}}} \right)^H}$\vspace{0.05cm}.

The EVD of the covariance matrix $\bm{\Psi}$ can be represented as 
\begin{equation}
\label{deqn_ex39}
\begin{array}{ll}
\boldsymbol \Psi\!\!\!\!\!\!&= {\mathbf{U}}\boldsymbol \Sigma {{\mathbf{U}}^H}\\
&= \left[ {{{\mathbf{U}}_s} \ {\rm{ }}{{\mathbf{U}}_n}} \right]\left[ {\begin{array}{*{20}{c}}
{{\boldsymbol \Sigma _s}}&{{{\mathbf{0}}_{K \times (N_L\!-\!K)}}}\vspace{0.1cm}\\
{{{\mathbf{0}}_{(N_L\!-\!K) \times K}}}&{\sigma_v^2{{\mathbf{I}}_{N_L\!-\!K}}}
\end{array}} \right]\left[ {\begin{array}{*{20}{c}}
{{\mathbf{U}}_s^H} \vspace{0.1cm}\\
{{\mathbf{U}}_n^H}
\end{array}} \right] \vspace{0.15cm}\\
&= {{\mathbf{U}}_s}{{\boldsymbol{\Sigma }}_s}{\mathbf{U}}_s^H + \sigma_v^2{{\mathbf{U}}_n}{\mathbf{U}}_n^H,
\end{array}
\end{equation}
where $\mathbf{U}$ is a ${N_L} \times {N_L}$ unitary matrix, $\boldsymbol \Sigma$ is a ${N_L} \times {N_L}$ diagnol matrix composed of the eigenvalues of $\boldsymbol \Psi$, $\boldsymbol \Sigma _s$ is a $K \times K$ diagnol matrix consisting of the $K$ largest eigenvalues of $\boldsymbol \Psi$, ${\mathbf{U}}_s$ is a $N_L \times K$ matrix which denotes the signal subspace, and ${\mathbf{U}}_n$ is a $N_L \times (N_L-K)$ matrix  denoting the noise subspace.

Notice that ${\mathbf{U}}=[ {{{\mathbf{U}}_s}\ {{\mathbf{U}}_n}} ]$ is a unitary matrix with the equality ${\mathbf U}\mathbf U^H={\mathbf U_s}\mathbf U_s^H + {\mathbf U_n}\mathbf U_n^H = {\mathbf I_{{N_L}}}$. Substituting ${\mathbf I_{{N_L}}}={\mathbf U_s}\mathbf U_s^H + {\mathbf U_n}\mathbf U_n^H$ into the right hand side of (\ref{deqn_ex38}) and
from (\ref{deqn_ex39}), the following equality holds
\begin{equation}
\label{deqn_ex40}
{\tilde{\mathbf A}}\boldsymbol \Gamma {\tilde{\mathbf A}}^H + {\sigma_v^2}{\mathbf U_s}\mathbf U_s^H = {\mathbf U_s}\boldsymbol \Sigma_s \mathbf U_s^H.
\end{equation}

Postmultiplying both sides of (\ref{deqn_ex40}) by $\mathbf U_s$, and according to $\mathbf U_s^H{\mathbf U_s} = {\mathbf I_K}$, it holds that   
\begin{equation}
\label{deqn_ex41}
{\tilde{\mathbf A}}\boldsymbol \Gamma {\tilde{\mathbf A}}^H{\mathbf U_s} + {\sigma_v^2}{\mathbf U_s}= {\mathbf U_s}\boldsymbol \Sigma_s.
\end{equation}

From (\ref{deqn_ex41}), we can obtain 
\begin{equation}
\label{deqn_ex42}
\begin{array}{ll}
{\mathbf U_s} \!\!\!\!\!\!&= {\tilde{\mathbf A}}\boldsymbol \Gamma {\tilde{\mathbf A}}^H{\mathbf U_s}{({\boldsymbol \Sigma _s} - {\sigma_v^2}{\mathbf I_K})^{ - 1}}\vspace{0.15cm}\\
 &= {\tilde{\mathbf A}}\mathbf H,
\end{array}
\end{equation}
where \vspace{0.05cm} $\mathbf H = \boldsymbol \Gamma {\tilde{\mathbf A}}^H{\mathbf U_s}{({\boldsymbol \Sigma _s} - {\sigma_v^2}{\mathbf I_K})^{ \!-\! 1}}\in \mathbb{C}^{K \times K}$. Since $K = {\rm{rank}}({\mathbf U_s}) = {\rm{rank}}({\tilde{\mathbf A}}\mathbf H) \le {\rm{rank}}(\mathbf H)$ and \vspace{0.05cm} ${\rm{rank}}(\mathbf H) \le K$, it holds that ${\rm{rank}}(\mathbf H) = K$, i.e., $\mathbf H$ is a full-rank square matrix.

It is noticed that the computational complexity for obtaining the signal subspace $\mathbf U_s$ \vspace{0.05cm} by performing EVD on the covariance $\bm \Psi$ is ${O}((\tilde M \!\tilde N \!\tilde P)^3)$, which leads to high computational cost, especially when $\tilde M\! \tilde N\! \tilde P$ is large. The FSD method \cite{ref42} in Algorithm 1 can be used to obtain $\mathbf U_s$ with the computational complexity $O(d(\tilde M \!\tilde N \!\tilde P)^2)$, where $d$ is the number of FSD iterations and greatly smaller than $n$. 
\begin{algorithm}[!t] 
\caption{FSD Algorithm.} 
\label{algorithm 2 caption} 
\begin{algorithmic}[1] %这个1 表示每一行都显示数字
\REQUIRE  %算法的输入参数：Input
$K,\boldsymbol{\Psi}$.\\
\STATE Construct a unit norm vector $\mathbf r_0$, and let ${b_0} = 1, {\mathbf q_0} = \mathbf 0 $.
\FOR{$j = 1, \ldots ,d$} 
\STATE \ ${\mathbf q_j} = {{{\mathbf r_{j - 1}}} \mathord{\left/{\vphantom {{{r_{j - 1}}} {{b_{j - 1}}}}} \right.\kern-\nulldelimiterspace} {{b_{j - 1}}}}$ \vspace{0.1cm}
\STATE \ ${ a_j} = \mathbf q_j^H\boldsymbol{ \Psi} {\mathbf q_j}$ \vspace{0.1cm}
\STATE \ ${\mathbf r_j} =\boldsymbol{ \Psi} {\mathbf q_j} - {a_j}{\mathbf q_j} - {b_{j - 1}}{\mathbf q_{j - 1}}$ \vspace{0.1cm}
\STATE \ ${b_j} = {\left\| {{\mathbf r_j}} \right\|_2}$ \vspace{0.1cm}
\ENDFOR
\label{ algorithm 2 step 1 }%对此行的标记，方便在文中引用算法的某个步骤
\STATE Define $\mathbf G \buildrel \Delta \over = \left[ {{\mathbf q_1}, \ldots ,{\mathbf q_d}} \right]$ and calculate the EVD of ${\mathbf G^H}\boldsymbol{\Psi} {\mathbf G}$ to obtain matrix $\mathbf E=[{\mathbf{ e}_0},\ldots ,{\mathbf{e}_{K \!-\!1}}]$, which is composed of the $K$ eigenvectors related to the largest $K$ eigenvalues of ${\mathbf G^H}\boldsymbol{\Psi} {\mathbf G}$.
\STATE Calculate ${\mathbf{ U}_s} = \mathbf G \mathbf E$.
\ENSURE ${\mathbf{U}_s}$.
\end{algorithmic}
\end{algorithm}

In order to utilize the translational invariance of the signal subspace $\mathbf U_s$, the following matrices are constructed 
\begin{align}
\label{deqn_ex43}
&{\mathbf{J}}_1^R \buildrel \Delta \over = {{\mathbf{I}}_{\tilde P}} \otimes \left[ {{{\mathbf{I}}_{\tilde M(\tilde N - 1)}},{{\mathbf{0}}_{\tilde M(\tilde N - 1) \times \tilde M}}} \right],  \\
\label{deqn_ex44}
&{\mathbf{J}}_2^R \buildrel \Delta \over = {{\mathbf{I}}_{\tilde P}} \otimes \left[ {{{\mathbf{0}}_{\tilde M(\tilde N - 1) \times \tilde M}},{{\mathbf{I}}_{\tilde M(\tilde N - 1)}}} \right],  \\
\label{deqn_ex45}
&{\mathbf{J}}_1^v \buildrel \Delta \over = {{\mathbf{I}}_{\tilde P}} \otimes {{\mathbf{I}}_{\tilde N}} \otimes \left[ {{{\mathbf{I}}_{(\tilde M - 1)}},{{\mathbf{0}}_{(\tilde M - 1) \times 1}}} \right],\\
\label{deqn_ex46}
&{\mathbf{J}}_2^v \buildrel \Delta \over = {{\mathbf{I}}_{\tilde P}} \otimes {{\mathbf{I}}_{\tilde N}} \otimes \left[ {{{\bf{0}}_{(\tilde M - 1) \times 1}},{{\mathbf{I}}_{(\tilde M - 1)}}} \right], \\
\label{deqn_ex47}
&{\mathbf{J}}_1^\theta  \buildrel \Delta \over = \left[ {{{\mathbf{I}}_{\tilde M\tilde N(\tilde P - 1)}},{{\bf{0}}_{\tilde M\tilde N(\tilde P - 1) \times \tilde M\tilde N}}} \right],\\
\label{deqn_ex48}
&{\mathbf{J}}_2^\theta  \buildrel \Delta \over = \left[ {{{\bf{0}}_{\tilde M\tilde N(\tilde P - 1) \times \tilde M\tilde N}},{{\mathbf{I}}_{\tilde M\tilde N(\tilde P - 1)}}} \right].
\end{align}	

Then, we have  
\begin{align}
\label{deqn_ex49}
&{\mathbf{J}_1^R{\tilde{\mathbf A}}} = \left [ {\tilde{\mathbf{a}}_0}(\!\tilde P,\!\tilde M,\!\tilde N\!\!-\!\!1\!), \ldots, {\tilde{\mathbf{a}}_{K\!-\!1\!}}(\!\tilde P,\!\tilde M,\!\tilde N\!\!-\!\!1\!) \right],\\
\label{deqn_ex50}
&{\mathbf{J}_2^R{\tilde{\mathbf A}}} = \left [ {\tilde{\mathbf{a}}_0}(\!\tilde P,\!\tilde M,\!\tilde N\!\!-\!\!1\!), \ldots, {\tilde{\mathbf{a}}_{K\!-\!1\!}}(\!\tilde P,\!\tilde M,\!\tilde N\!\!-\!\!1\!)\right ]\!{\mathbf{\Lambda}\!^R},\\
\label{deqn_ex51}
&{\mathbf{J}_1^v{\tilde{\mathbf A}}} = \left [ {\tilde{\mathbf{a}}_0}(\!\tilde P,\!\tilde M\!\!-\!\!1,\!\tilde N\!), \ldots, {\tilde{\mathbf{a}}_{K\!-\!1\!}}(\!\tilde P,\!\tilde M\!\!-\!\!1,\!\tilde N\!) \right],\\
\label{deqn_ex52}
&{\mathbf{J}_2^v{\tilde{\mathbf A}}} = \left [ {\tilde{\mathbf{a}}_0}(\!\tilde P,\!\tilde M\!\!-\!\!1,\!\tilde N\!), \ldots, {\tilde{\mathbf{a}}_{K\!-\!1\!}}(\!\tilde P,\!\tilde M\!\!-\!\!1,\!\tilde N\!)\right ]\!{\boldsymbol{\Lambda}\!^v},\\
\label{deqn_ex53}
&{\mathbf{J}_1^\theta{\tilde{\mathbf A}}} = \left [ {\tilde{\mathbf{a}}_0}(\!\tilde P\!\!-\!\!1,\!\tilde M,\!\tilde N\!), \ldots, {\tilde{\mathbf{a}}_{K\!-\!1\!}}(\!\tilde P\!\!-\!\!1,\!\tilde M,\!\tilde N\!) \right],\\
\label{deqn_ex54}
&{\mathbf{J}_2^\theta{\tilde{\mathbf A}}} = \left [ {\tilde{\mathbf{a}}_0}(\!\tilde P\!\!-\!\!1,\!\tilde M,\!\tilde N\!), \ldots, {\tilde{\mathbf{a}}_{K\!-\!1\!}}(\!\tilde P\!\!-\!\!1,\!\tilde M,\!\tilde N\!)\right ]\!{\boldsymbol{\Lambda}\!^\theta},
\end{align}
where \vspace{0.05cm} ${{\mathbf{\Lambda}}\!^\theta} \!\!= \!\!{\rm{diag}}\!\left\{\! {{e^{j {\psi _0}}},\ldots,{e^{j{\psi_{\!K\!-\!1}}}}}\!\!\right\}, {{\boldsymbol{\Lambda}}\!^v}\!\! =\!\! {\rm{diag}}\!\left\{ {{e^{j {\phi_0}}},\!\ldots\!,{e^{j {\phi_{\!K\!-\!1}}}}}\!\! \right\}$, and ${{\boldsymbol{\Lambda}}\!^R}\!\!=\!\!{\rm{diag}}\!\left\{ {{e^{ \!-\! j {\varphi_0}}},\ldots,{e^{ \!-\!j {\varphi_{\!K\!-\!1}}}}}\!\!\right\}$. \vspace{0.05cm} Suppose $\tilde P$, $\tilde M$ and $\tilde N$ are large enough such that ${\mathbf{J}_1^R{\tilde{\mathbf A}}}$, ${\mathbf{J}_1^v{\tilde{\mathbf A}}}$ and ${\mathbf{J}_1^\theta{\tilde{\mathbf A}}}$ are all column full-rank matrices.

Furthermore, both sides of (\ref{deqn_ex42}) are left multiplied by the matrices shown in (\ref{deqn_ex43})-(\ref{deqn_ex48}), respectively, and we can that 
\begin{align}
\label{deqn_ex55}
&{\mathbf{U}}_1^R \buildrel \Delta \over = {\mathbf{J}}_1^R{{\mathbf{U}}_s} = {\mathbf{J}}_1^R{{{\tilde{\mathbf A}}{\mathbf H}}},\\
\label{deqn_ex56}
&{\mathbf{U}}_2^R \buildrel \Delta \over = {\mathbf{J}}_2^R{{\mathbf{U}}_s} = {\mathbf{J}}_2^R{{\tilde{\mathbf A}}{\mathbf H}} = {\mathbf{J}}_1^R{{\tilde{\mathbf A}}}{{\boldsymbol{\Lambda }}\!^R}{{{\mathbf H}}},\\
\label{deqn_ex57}
&{\mathbf{U}}_1^v \buildrel \Delta \over = {\mathbf{J}}_1^v{{\mathbf{U}}_s} = {\mathbf{J}}_1^v{{\tilde{\mathbf A}}{\mathbf H}},\\
\label{deqn_ex58}
&{\mathbf{U}}_2^v \buildrel \Delta \over = {\mathbf{J}}_2^v{{\mathbf{U}}_s} = {\mathbf{J}}_2^v{{\tilde{\mathbf A}}{\mathbf H}} = {\mathbf{J}}_1^v{{\tilde{\mathbf A}}}{{\boldsymbol{\Lambda }}\!^v}{\mathbf{{H}}},\\
\label{deqn_ex59}
&{\mathbf{U}}_1^\theta  \buildrel \Delta \over = {\mathbf{J}}_1^\theta {{\mathbf{U}}_s} = {\mathbf{J}}_1^\theta {{\tilde{\mathbf A}}{\mathbf H}},\\
\label{deqn_ex60}
&{\mathbf{U}}_2^\theta  \buildrel \Delta \over = {\mathbf{J}}_2^\theta {{\mathbf{U}}_s} = {\mathbf{J}}_2^\theta {{\tilde{\mathbf A}}{\mathbf H}} = {\mathbf{J}}_1^\theta {{\tilde{\mathbf A}}}{{\mathbf{\Lambda }}\!^\theta }{\mathbf{{\mathbf H}}}.
\end{align}	

Since $\mathbf H$ is a full-rank matrix, substituting (\ref{deqn_ex55}), (\ref{deqn_ex57}) and (\ref{deqn_ex59}) into (\ref{deqn_ex56}), (\ref{deqn_ex58}) and (\ref{deqn_ex60}), respectively, yields
\begin{align}
\label{deqn_ex61}
&{\mathbf{U}}_2^R{\rm{ = }}{\mathbf{U}}_1^R{{\mathbf{{ H}}}^{ \!-\! 1}}{{\mathbf{\Lambda}}\!^R}{\mathbf{{H}}},\\
\label{deqn_ex62}
&{\mathbf{U}}_2^v{\rm{ = }}{\mathbf{U}}_1^v{{\mathbf{{ H}}}^{ \!-\! 1}}{{\mathbf{\Lambda}}\!^v}{\mathbf{{H}}},\\
\label{deqn_ex63}
&{\mathbf{U}}_2^\theta {\rm{ = }}{\mathbf{U}}_1^\theta {{\mathbf{{ H}}}^{ \!-\! 1}}{{\mathbf{\Lambda}}\!^\theta }{\mathbf{{H}}}.
\end{align}	

Since ${\mathbf{J}_1^R{\tilde{\mathbf A}}}$, ${\mathbf{J}_1^v{\tilde{\mathbf A}}}$ and ${\mathbf{J}_1^\theta{\tilde{\mathbf A}}}$ are all column full-rank matrices and $\mathbf H$ is a full-rank square matrix, the ${\mathbf{U}}_1^R$, ${\mathbf{U}}_1^v$ and ${\mathbf{U}}_1^\theta$ are all column full-rank matrices. Then, from (\ref{deqn_ex61}), (\ref{deqn_ex62}) and (\ref{deqn_ex63}), the following equalities hold 
\begin{align}
\label{deqn_ex64}
&{{\mathbf{T}}^R} \buildrel \Delta \over = {{\mathbf{{H}}}^{ \!-\! 1}}{{\mathbf{\Lambda}}\!^R}{\mathbf{{H}}}
= {({\mathbf{U}}_1^R)^{\dag}}{\mathbf{U}}_2^R,\\
\label{deqn_ex65}
&{{\mathbf{T}}^v} \buildrel \Delta \over = {{\mathbf{{ H}}}^{ \!-\! 1}}{{\boldsymbol{\Lambda}}\!^v}{\mathbf{{H}}} = {({\mathbf{U}}_1^v)^{\dag}}{\mathbf{U}}_2^v ,\\
\label{deqn_ex66}
&{{\mathbf{T}}^\theta } \buildrel \Delta \over  = {{\mathbf{{H}}}^{ \!-\! 1}}{{\boldsymbol{\Lambda}}\!^\theta }{\mathbf{{H}}} = {({\mathbf{U}}_1^\theta )^{\dag}}{\mathbf{U}}_2^\theta .
\end{align}	

According to (\ref{deqn_ex64}), it can be checked that the eigenvalues of ${\mathbf{T}}^R$ are the diagonal elements of ${{\boldsymbol{\Lambda}}\!^R}$, and the corresponding eigenvectors are the columns of  ${{\mathbf{{H}}}^{\!-\!1}}$. Hence, the elements of ${{\boldsymbol{\Lambda}}\!^R}$ can be obtained by employing the EVD on ${\mathbf{T}}^R$, then, the ranges of targets can be calculated from these eigenvalues. Similarly, the diagonal elements of ${{\boldsymbol{\Lambda}}\!^v}$ and ${{\boldsymbol{\Lambda}}\!^\theta}$ can be obtained by performing the EVD on ${\mathbf{T}}^v$ and ${\mathbf{T}}^\theta$, respectively. Then, the velocities and azimuths of targets can be obtained, respectively. 

As shown in (\ref{deqn_ex64}), (\ref{deqn_ex65}) and (\ref{deqn_ex66}), ${{\mathbf{T}}^R}$ is determined by ${\mathbf{U}}_1^R$ and ${\mathbf{U}}_2^R$, ${{\mathbf{T}}^v}$ depends on ${\mathbf{U}}_1^v$ and ${\mathbf{U}}_2^v$, and ${{\mathbf{T}}^\theta }$ is decided by ${\mathbf{U}}_1^\theta$ and ${\mathbf{U}}_2^\theta$. In addition, all the ${\mathbf{U}}_1^R$, ${\mathbf{U}}_2^R$, ${\mathbf{U}}_1^v$, ${\mathbf{U}}_2^v$, ${\mathbf{U}}_1^\theta$ and ${\mathbf{U}}_2^\theta$ are determined by the covariance matrix $\bm{\Psi}$. In brief, ${{\mathbf{T}}^R}$, ${{\mathbf{T}}^v}$, and ${{\mathbf{T}}^\theta }$ are all determined by $\bm{\Psi}$. In practice, the statistical covariance matrix $\bm{\Psi}$ is unavailable, and the observation signal in (\ref{deqn_ex36}) is usually used to calculate the sample covariance matrix $\bm{\hat \Psi}$. The sample covariance matrix $\bm{\hat \Psi}$ can be expressed as 

\begin{equation}
\label{deqn_ex67}
\mathbf{\hat \Psi} {\rm{ = }}\frac{1}{{{N_{{\rm{s}}}}}}\sum\limits_{\tilde p = 0}^{P \!-\! \tilde P} {\sum\limits_{\tilde m = 0}^{M \!-\!\tilde M} {\sum\limits_{\tilde n = 0}^{N \!-\!\tilde N} {{\mathbf g_{\tilde p,\tilde m,\tilde n}}} } } \mathbf g_{\tilde p,\tilde m,\tilde n}^H.
\end{equation}

Using the sample covariance matrix $\boldsymbol{\hat{\Psi}}$ in (\ref{deqn_ex67}), we can obtain the estimators
${{\mathbf{\hat{T}}}^R}$, ${{\mathbf{\hat{T}}}^v}$ and ${{\mathbf{\hat{T}}}^\theta }$. Suppose that the EVD of ${\mathbf{\hat T}^R}$ is expressed as ${\mathbf{\hat T}^R} = {\mathbf Q^{ \!-\! 1}}{\boldsymbol \Xi ^R}\mathbf Q$. Then, we have 
\begin{equation}
\label{deqn_ex68}
{ \boldsymbol{\Xi}^R} = \mathbf Q{{\mathbf{\hat T}}^R}{\mathbf Q^{ \!-\! 1}},
\end{equation}
where $\mathbf Q$ is a matrix composed of all eigenvectors of $\mathbf{\hat T}^R$, and ${\bm \Xi ^R} = {\rm{diag}}\left\{ {\lambda _0^R,\lambda _1^R, \ldots ,\lambda _{K\!-\!1}^R} \right\}$ is a $K \times K$ diagonal matrix that collects all the eigenvalues of $\mathbf{\hat T}^R$.

From (\ref{deqn_ex68}), we obtain the range estimations as follows
\begin{equation}
\label{deqn_ex69}
{\hat R_i} =  - \frac{{{\rm{angle}}(\lambda _i^R) c}}{{4\pi \Delta f}}, \ {\rm for} \ i = 0,1,\ldots,K\!-\!1.
\end{equation}

As in (\ref{deqn_ex64}), (\ref{deqn_ex65}) and (\ref{deqn_ex66}), since ${{\mathbf{T}}^R}$, ${{\mathbf{T}}^v}$, and ${{\mathbf{T}}^\theta }$ have the same eigenvectors, their sample counterparts ${{\mathbf{\hat{T}}}^R}$, ${{\mathbf{\hat{T}}}^v}$, and ${{\mathbf{\hat{T}}}^\theta }$  have the same eigenvectors in a statistic sense. Hence, we can use the eigenvector matrix $\mathbf Q$ in (\ref{deqn_ex68}) to obtain the eigenvalue estimates of ${{\mathbf{\hat{T}}}^v}$ and ${{\mathbf{\hat{T}}}^\theta }$, i.e.,
\begin{equation}
\label{deqn_ex70}
{\rm Diag}\{ \mathbf Q{\mathbf{\hat T}^v}{\mathbf Q^{ \!-\! 1}}\}=\boldsymbol{\Xi}^v,
\end{equation}
\begin{equation}
\label{deqn_ex71}
{\rm Diag}\{ \mathbf Q{\mathbf{\hat T}^\theta}{\mathbf Q^{ \!-\! 1}}\}=\boldsymbol{\Xi}^\theta,
\end{equation}
where ${\bm \Xi ^v} = {\rm{diag}}\left\{ {\lambda _0^v,\ldots,\lambda_{K\!-\!1}^v} \right\}$ is a diagonal matrix whose diagonal elements are corresponding to the diagonal elements of $\mathbf Q{\mathbf{\hat T}^v}{\mathbf Q^{ \!-\!1}}$. Similarly, ${\bm \Xi ^\theta} = {\rm{diag}}\left\{ {\lambda _0^\theta,\ldots,\lambda_{K\!-\!1}^\theta} \right\}$ is a diagonal matrix whose diagonal elements are corresponding to the diagonal elements of $\mathbf Q{\mathbf{\hat T}^\theta}{\mathbf Q^{ \!-\! 1}}$.
Furthermore, we can obtain the velocity and azimuth estimations as follows:
\begin{equation}
\label{deqn_ex72}
{\hat v_i} = \frac{{{\rm{angle(}}\lambda _i^v{\rm{)}}c}}{{4\pi {f_c}\bar T}},\ {\rm for} \ i = 0,1,\ldots,K\!-\!1,
\end{equation}
\begin{equation}
\label{deqn_ex73}
{\hat \theta _i} = \arcsin \left[ {\frac{{\lambda {\rm{angle(}}\lambda _i^\theta {\rm{)}}}}{{2\pi d}}} \right],\ {\rm for} \ i = 0,1,\ldots,K\!-\!1.
\end{equation}

Since $\boldsymbol \Xi^R$, $\boldsymbol \Xi^v$ and $\boldsymbol \Xi^\theta$ are obtained by using the same eigenvectors matrix $\mathbf Q$, the range estimations in (\ref{deqn_ex69}), the velocity estimations in (\ref{deqn_ex72}), and the azimuth estimations in (\ref{deqn_ex73}) are already paired, i.e., the ${\hat R_i}$ in (\ref{deqn_ex69}), the ${\hat v_i}$ in (\ref{deqn_ex72}), and the ${\hat \theta_i}$ in (\ref{deqn_ex73}) are the range, velocity, and azimuth estimations of the $i$-th target, respectively.

\subsection{Overall 3DJE Algorithm and Complexity Analysis}
The steps of the 3DJE are summarized in the following Algorithm 2.
\begin{algorithm}[!h]
\caption{3DJE Algorithm.} 
\label{alg:Framwork} 
\begin{algorithmic}[1] %这个1 表示每一行都显示数字
\REQUIRE  %算法的输入参数：Input
$K,M,N,P,\tilde M,\tilde N, \tilde P, \mathbf{b}$.\\
\STATE Perform smooth operation to obtain ${\mathbf g}_{\tilde p,\tilde m,\tilde n}$ in (\ref{deqn_ex34}).
\label{code:fram:trainbase}
\STATE Calculate the sample covariance matrix $\boldsymbol{\hat \Psi}$ by (\ref{deqn_ex67}).
\label{code:fram:add}
\STATE  Calculate the signal subspace estimation ${\mathbf{\hat U}}_s$ from the sample covariance matrix $\boldsymbol{\hat \Psi}$ by Algorithm 1.
\STATE Construct ${\mathbf{J}}_1^R$, ${\mathbf{J}}_2^R$,  ${\mathbf{J}}_1^v$,  ${\mathbf{J}}_2^v$, ${\mathbf{J}}_1^\theta$,  ${\mathbf{J}}_2^\theta$ in (\ref{deqn_ex43})-(\ref{deqn_ex48}). Then, calculate the  ${\mathbf{\hat  U}}_1^R={\mathbf{J}}_1^R{{\mathbf{\hat U}}_s}$, \vspace{0.1cm}${\mathbf{\hat U}}_2^R={\mathbf{J}}_2^R{{\mathbf{\hat U}}_s}$, ${\mathbf{\hat U}}_1^v={\mathbf{J}}_1^v{{\mathbf{\hat U}}_s}$, ${\mathbf{\hat U}}_2^v={\mathbf{J}}_2^v{{\mathbf{\hat U}}_s}$, ${\mathbf{\hat U}}_1^\theta={\mathbf{J}}_1^\theta{{\mathbf{\hat U}}_s}$, ${\mathbf{\hat U}}_2^\theta={\mathbf{J}}_2^\theta{{\mathbf{\hat U}}_s}$.\vspace{0.1cm}
\label{code:fram:select}
\STATE Calculate \vspace{0.05cm} ${{\mathbf{\hat T}}^R}$, ${\mathbf{\hat T}^v}$ and ${{\mathbf{\hat T}}^\theta}$ using (\ref{deqn_ex64}), (\ref{deqn_ex65}) and (\ref{deqn_ex66}), respectively.
\STATE Calculate the EVD of ${\mathbf{\hat T}}^R$ to obtain diagonal matrix ${\boldsymbol \Xi ^R} = \mathbf Q{{\mathbf{\hat T}}^R}{\mathbf Q^{ \!-\! 1}}$.
\STATE Calculate the range estimations ${\hat R_i}$ by (\ref{deqn_ex69}).
\STATE Calculate $\boldsymbol \Xi^v$ and $\boldsymbol \Xi^\theta$ by (\ref{deqn_ex70}) and (\ref{deqn_ex71}), respectively.
\STATE Calculate the velocity estimations ${\hat v_i}$ and azimuth estimations ${\hat \theta _i}$ by (\ref{deqn_ex72}) and (\ref{deqn_ex73}), respectively.
\ENSURE $({\hat R_i},{\hat v_i},{\hat \theta_i})$, for $ i = 0,1, \ldots ,K-1.$
\end{algorithmic}
\end{algorithm}

The computational of Algorithm 2 is analyzed as follows. The total computational complexity of step 1 and step 2 is ${O}({N_{{\rm{s}}}}{N_L^2})$. \vspace{0.05cm} The computational complexity of  step 3 is ${O}(d{N_L^2})$.  The computational cost of step 4 is ${O}(2{N_L}{{\tilde P}}(\tilde M \!-\! 1){{\tilde N}}K \!+\! 2{N_L}{{\tilde P}}{{\tilde M}}(\tilde N \!-\!1)K \!+\! 2{N_L}(\tilde P \!-\! 1){{\tilde M}}{{\tilde N}}K)$. The computational complexity of step 5 is ${O}(3{K^3} \!+\! 3(\tilde P \!-\! 1)\tilde M\tilde N{K^2}\! +\! 3\tilde P(\tilde M \!- \!1)\tilde N {K^2} \!+\! 3\tilde P\tilde M(\tilde N \!-\! 1){K^2})$. The total computational cost of step 6 and step 7 is ${O}({K^3}+K)$. The total computational complexity of step 8 and step 9 is ${O}({4K^3}+4K)$. Therefore, the total computational complexity of Algorithm 2 is approximately \vspace{0.05cm} ${O}(({N_{{\rm{s}}}}+d){N_L^2}\!+\!({{\tilde P}\!-\!1}){\tilde M}{{\tilde N}}(2N_LK\!+\!3K^2) \!+\!{{\tilde P}}(\tilde M\!-\!1){{\tilde N}}(2N_LK\!+\!3K^2)+ {{\tilde P}}\tilde M({{\tilde N}\!-\!1})(2N_LK\!+\!3K^2)\!+\! 8{K^3})$.

\section{Simulation Results}
In this section, extensive simulation results are given to verify the performance of the proposed algorithm. We consider the parameter set of 5G NR, and the carrier frequency is ${f_c} =$ 27 GHz. The signal bandwidth is $B = $ 14.4 MHz and the CPI is 1 ms (i.e., the duration of each subframe). When $\Delta f =$ 60 kHz, there are 20 resource blocks (RBs), the number of subcarriers and OFDM symbol is $N=$ 240 and $M = $ 56, respectively. 
When $\Delta f = $ 120 kHz, there are 10 RBs, and the number of subcarriers and OFDM symbol is $N=$ 120 and  $M =$ 112, respectively. The number of receive antennas is $P=$ 8, and the inter-antenna spacing is $d=\lambda/2$. The detailed parameters are shown in Table \textrm{I}. In addition, the parameters of the smoothing operation are set as $\tilde M = $ 15, $\tilde N = $ 15 and $\tilde P =$ 7. The quadrature phase-shift keying (QPSK) modulation scheme is utilized to generate data symbol $s_m(n)$.
\begingroup
\begin{table}[!h]
\caption{System Parameters\label{tab:table1}}
\renewcommand\arraystretch{0.95}
\centering
\begin{tabular}{|m{0.8cm}<{\centering}|m{4.8cm}<{\centering}| m{1.8cm}<{\centering}|}
\hline 
{\textbf{Symbol}} & {\textbf{Parameter}} & {\textbf{Value}} \\
\hline
\makecell*[c]{$f_c$} & \makecell*[c]{Carrier frequency (GHz)}& \makecell*[c]{27} \\
\hline
\makecell*[c]{$\Delta f$} & \makecell*[c]{Subcarrier spacing (kHz)} & \makecell*[c]{60, 120} \\
\hline
\makecell*[c]{$M$} & \makecell*[c]{Number of OFDM symbols} & \makecell*[c]{56, 112}\\
\hline
\makecell*[c]{$N$} & \makecell*[c]{Number of subcarriers} & \makecell*[c]{240, 120}\\
\hline
\makecell*[c]{$B$} & \makecell*[c]{Bandwidth (MHz)} & \makecell*[c]{14.4}\\
\hline
\makecell*[c]{$T$} & \makecell *[c]{Elementary OFDM symbol duration (${\mu}$s)}  & \makecell*[c]{16.7, 8.33 }\\
\hline 
\makecell*[c]{$T_{\rm{cp}}$} & \makecell*[c]{Cyclic prefix duration (${\mu}$s)}   & \makecell*[c]{1.2, 0.59 }\\
\hline
\makecell*[c]{$\bar T$} & \makecell*[c]{Total OFDM symbol duration (${\mu}$s)} & \makecell*[c]{17.9, 8.92 }\\
\hline
\makecell*[c]{${R}_{\rm max}$} & \makecell*[c]{Maximum detectable range (m)}  & \makecell*[c]{180, 88.5 }\\
\hline
\makecell*[c]{$\Delta R$} & \makecell*[c]{Range resolution (m)}  & \makecell*[c]{10.42}\\
\hline
\makecell*[c]{$v_{\rm u}$} & \makecell*[c]{Maximum unambiguous velocity (m/s)}  & \makecell*[c]{$\pm \text{155.2}, \pm \text{311.4}$}\\
\hline
\makecell*[c]{$\Delta v$} & \makecell*[c]{Velocity resolution (m/s)}  & \makecell*[c]{5.54}\\
\hline
\makecell*[c]{$N_{\rm{Tx}}$} & \makecell*[c]{Number of transmit antennas} &  \makecell*[c]{1} \\
\hline
\makecell*[c]{$P$} & \makecell*[c]{Number of receive antennas} &  \makecell*[c]{8} \\
\hline
\end{tabular}
\end{table}

The RMSE is a common metric to compare the performance of different estimation methods and it is defined as:
\begin{equation}
\label{deqn_ex74}
{\rm{RMSE}} = \sqrt {\frac{{\left\| {\boldsymbol{\uprho}  -\boldsymbol{ \hat \uprho} } \right\|_2^2}}{K}},
\end{equation}
where $\boldsymbol{\uprho}$  denotes the true values of parameters, and $\boldsymbol{ \hat \uprho}$ denotes the estimated values. 
\subsection{Estimation Accuracy for Single Target}
We set the number of target $K=$ 1, and the range, velocity, and azimuth parameters of the single target are assumed to be $(35{\rm{m, 15m/s, 20}}^\circ )$. The SNR varies from -20 dB to 5 dB with an interval of 5 dB, and for each SNR, 200 times independent Monte Carlo simulations are performed. The RMSE simulation results are shown in Fig. \ref{fig3} and Fig. \ref{fig4}. 

\begin{figure}[!t]
\centering
\subfloat[RMSE of range estimation vs SNR.]{
\includegraphics[width=2.7in]{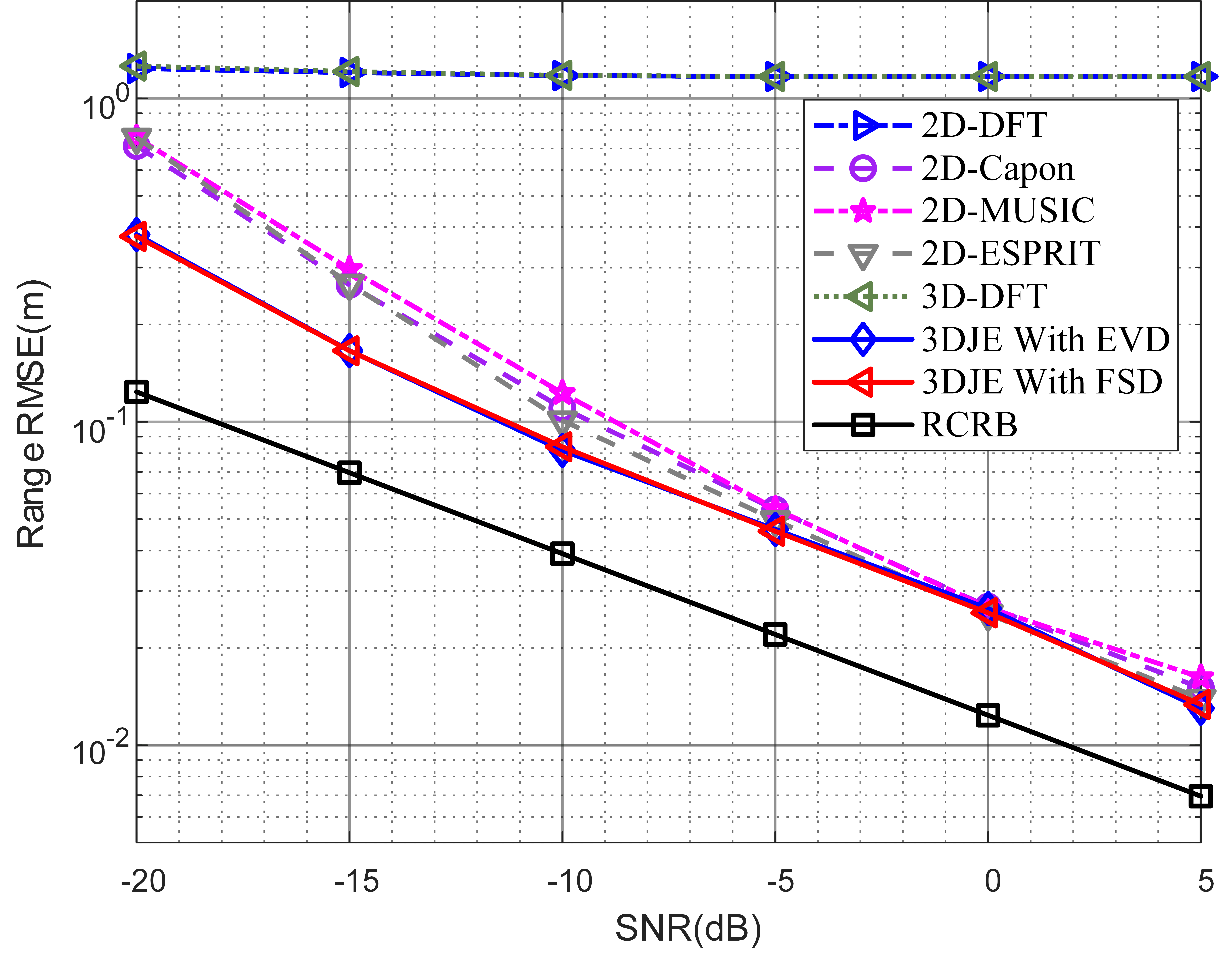} 
}

\subfloat[RMSE of velocity estimation vs SNR.]{
\includegraphics[width=2.7in]{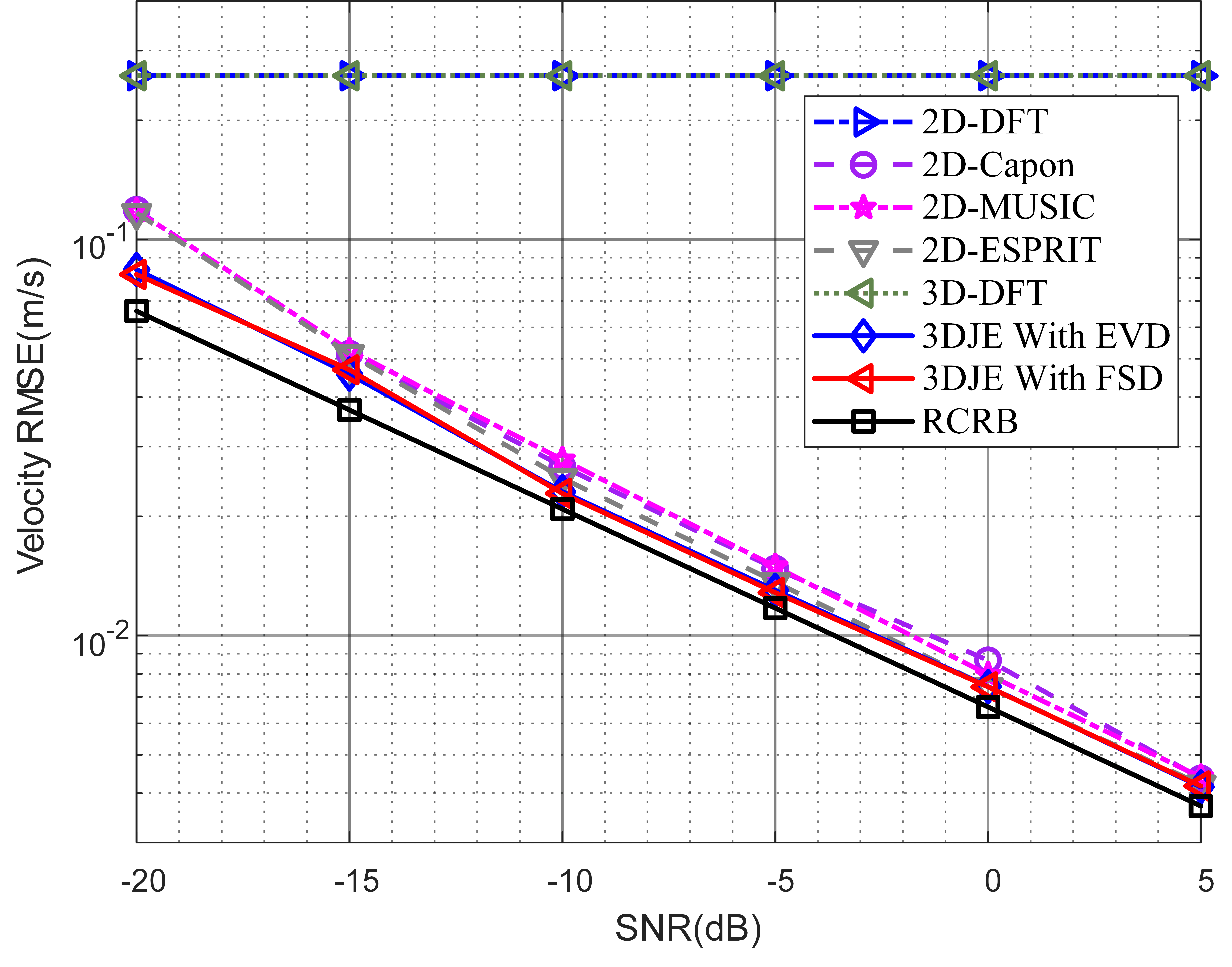} 
}

\subfloat[RMSE of azimuth estimation vs SNR.]{
\includegraphics[width=2.7in]{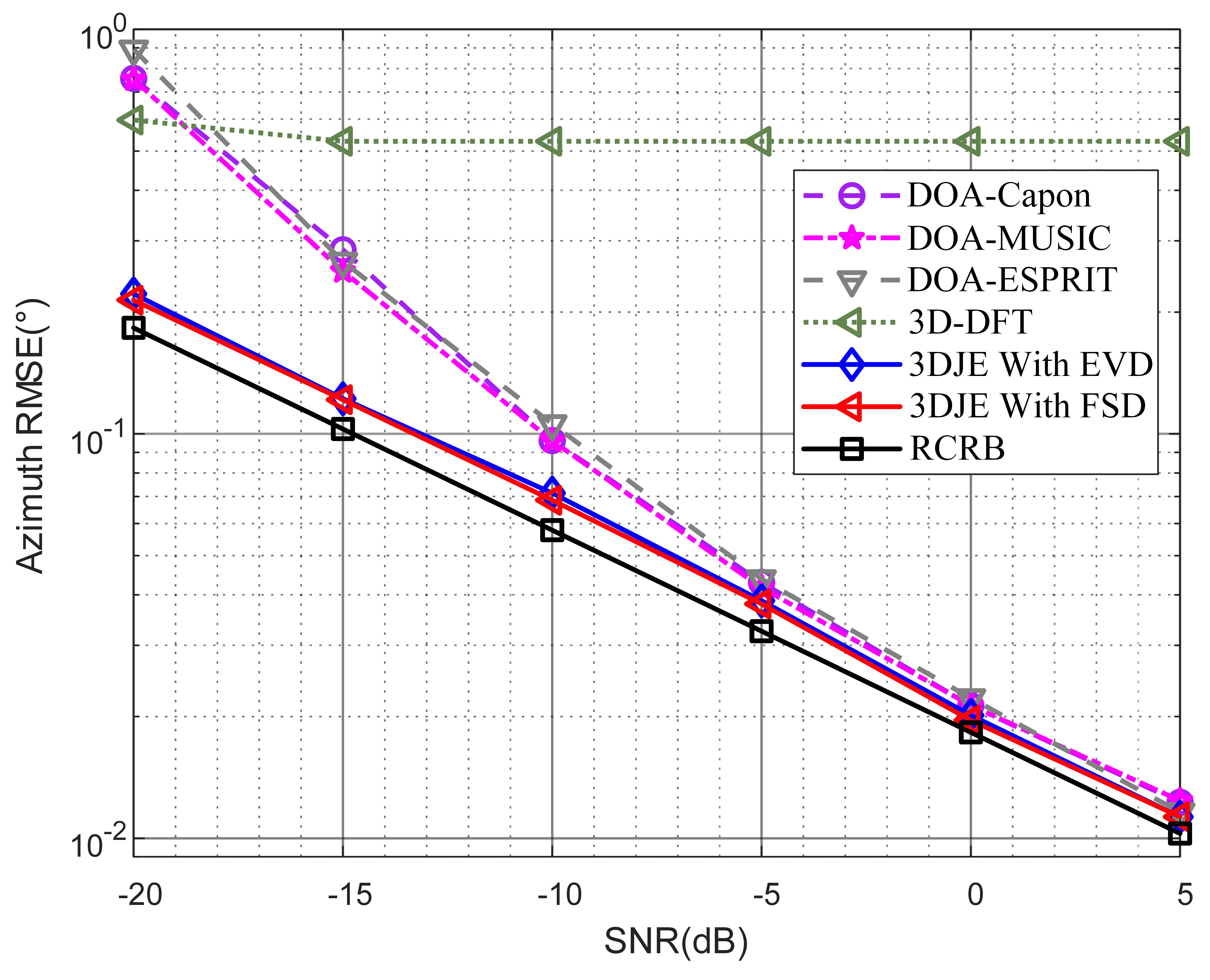} 
}

\caption{Single-target estimation RMSE vs SNR for $\Delta f = 60\ {\rm{ }}\rm kHz$.}
\label{fig3}
\end{figure}

\begin{figure}[!t]
\centering
\subfloat[RMSE of range estimation vs SNR.]{
\includegraphics[width=2.7in]{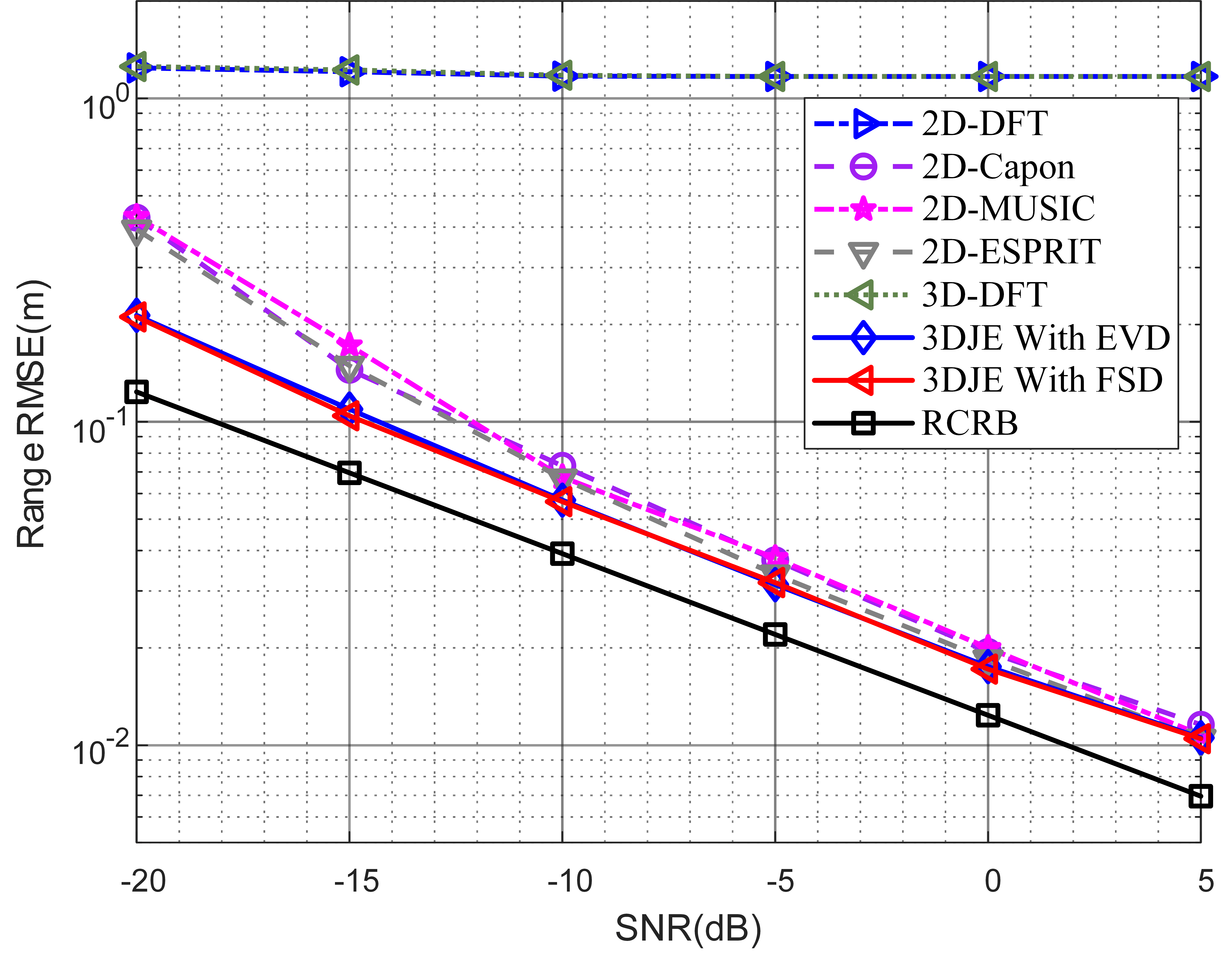} 
}

\subfloat[RMSE of velocity estimation vs SNR.]{
\includegraphics[width=2.7in]{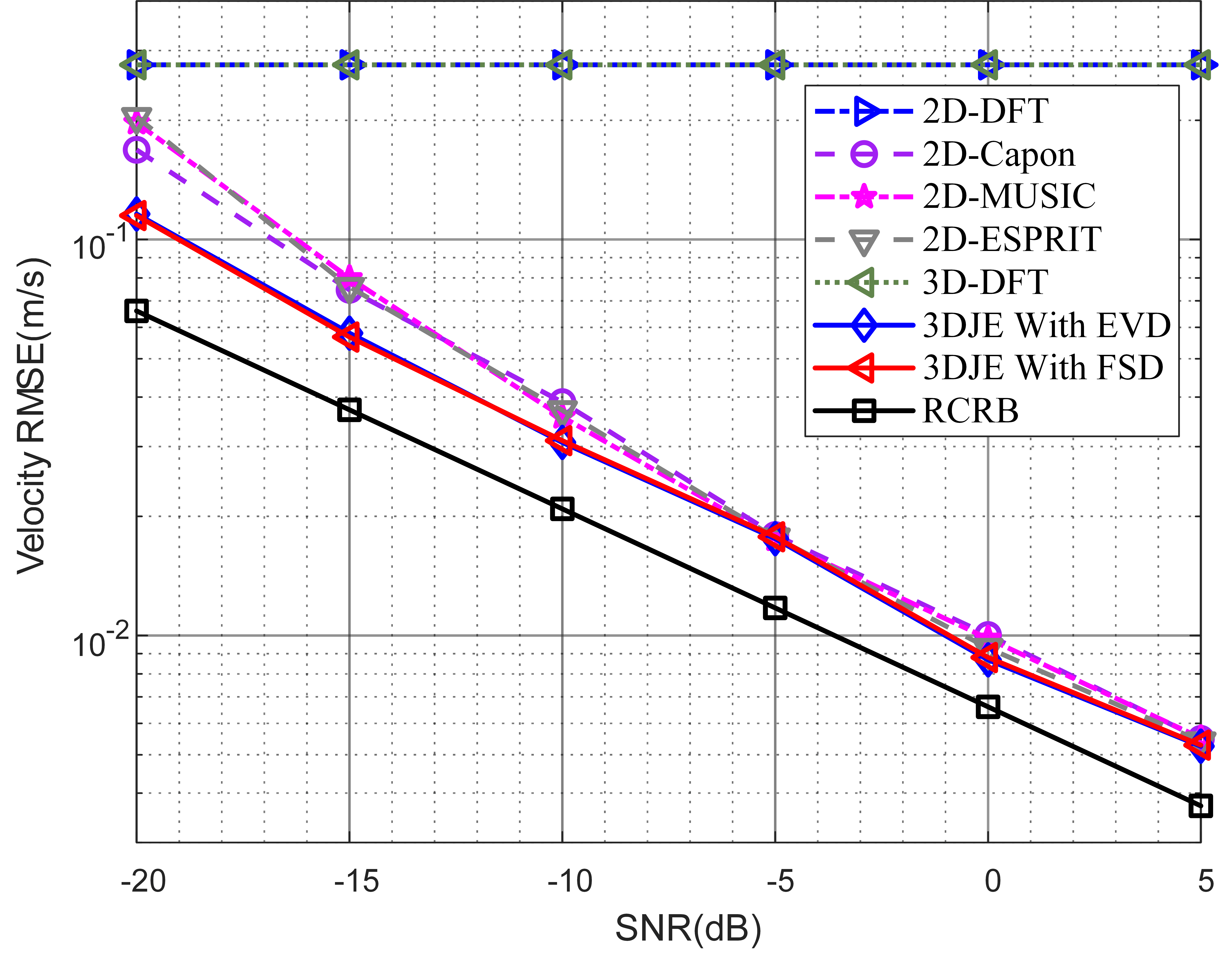} 
}

\subfloat[RMSE of azimuth estimation vs SNR.]{
\includegraphics[width=2.7in]{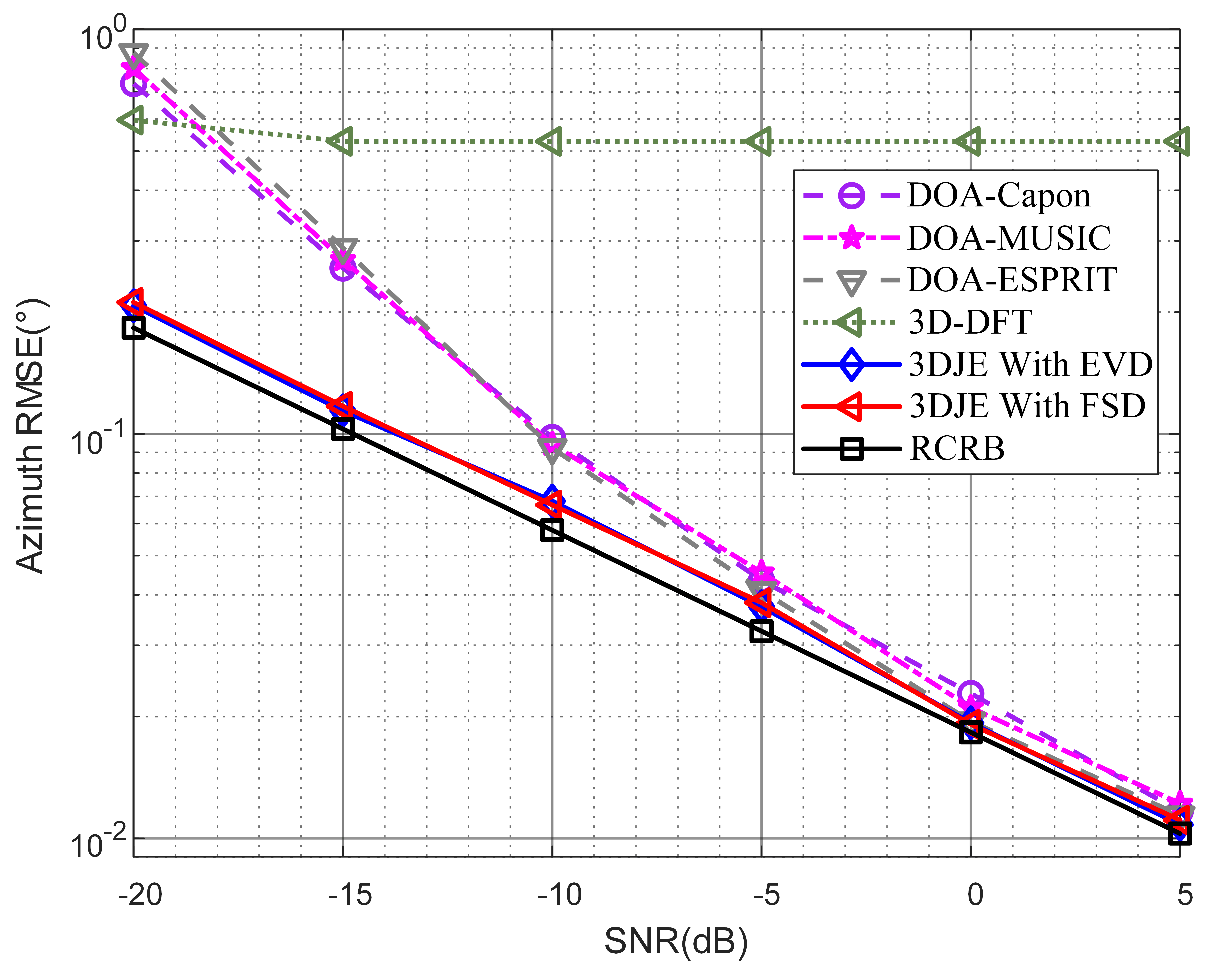} 
}

\caption{Single-target estimation RMSE vs SNR for $\Delta f = 120\ {\rm{ }}\rm kHz$.}
\label{fig4}
\end{figure}
As illustrated in Fig. \ref{fig3} for $\Delta f=$ 60 kHz and Fig. \ref{fig4} for $\Delta f= $ 120 kHz, respectively. The 3DJE with FSD and 3DJE with EVD denote signal subspace estimation obtained by FSD and EVD, respectively. The 2D-DFT, 2D-MUSIC, 2D-ESPRIT, 2D-Capon and 3D-DFT denote the aforementioned estimation methods which are detailed in \cite{ref12}, \cite{ref31}, \cite{ref32} and \cite{ref37}. The DOA-Capon, DOA-MUSIC, and DOA-ESPRIT in legend denote Capon \cite{ref34}, MUSIC \cite{ref35}, and ESPRIT \cite{ref36} methods for azimuth estimation, respectively. Numerical results show that the proposed 3DJE algorithm with FSD achieves almost the same RMSE performance as the 3DJE with EVD.

In Fig. \ref{fig3}(a) and Fig. \ref{fig4}(a), the RMSE of range estimation with different SNRs is illustrated. The proposed 3DJE algorithm achieves better estimation performance than other algorithms, especially when the SNR is low. Specifically, at a RMSE level of 10$^{-1}$, for both $\Delta f = $ 60 kHz and $\Delta f = $ 120 kHz, the proposed algorithm achieves a SNR gain of 2.5 dB compared to the 2D-MUSIC method. Besides, for $\Delta f = $ 120 kHz, the proposed algorithm suffers from about 3.5 dB SNR degradation compared to the RCRB. It is noted that the RMSE of the traditional DFT-based methods (i.e., 2D-DFT and 3D-DFT methods) does not change with the increase of SNR, the reason is that the DFT-based methods can not achieve super-resolution estimation of range, velocity and azimuth, and the range resolution is determined by the bandwidth, that is, $\Delta R={c \mathord{\left/{\vphantom {c {2N\Delta f}}} \right.\kern-\nulldelimiterspace} {(2B)}}$. Moreover, the proposed algorithm achieves better range estimation performance for $\Delta f = $ 120 kHz than that for $\Delta f = $ 60 kHz.

In Fig. \ref{fig3}(b) and Fig. \ref{fig4}(b), the RMSE of velocity estimation with different SNRs is presented. Fig. \ref{fig3}(b) and Fig. \ref{fig4}(b) show that the proposed 3DJE algorithm achieves better estimation performance than other algorithms, especially when the SNR is low. Specifically, at a RMSE level of 10$^{-2}$, compared to the RCRB, the proposed algorithm suffers from only 0.9 dB and 2.7 dB SNR degradation for $\Delta f = $ 60 kHz and $\Delta f = $ 120 kHz, respectively. Since the velocity resolution of DFT-based methods is limited by the CPI, that is, $\Delta v ={c\mathord{\left/{\vphantom {c {2{f_c}M\bar T}}} \right.\kern-\nulldelimiterspace} {(2{f_c}M\bar T)}}$, the velocity RMSE of DFT-based methods remain constant. Moreover, the proposed algorithm achieves better velocity estimation performance for $\Delta f = $ 60 kHz than that for $\Delta f = $ 120 kHz.

In Fig. \ref{fig3}(c) and Fig. \ref{fig4}(c), the RMSE of azimuth estimation with different SNRs is demonstrated. Fig. \ref{fig3}(c) and Fig. \ref{fig4}(c) demonstrate that the proposed 3DJE algorithm achieves better estimation performance than other algorithms, especially when the SNR is low. Specifically, at a RMSE level of 10$^{-1}$, compared to the DOA-MUSIC method, the proposed algorithm achieves a SNR gain of 3.1 dB and 3.3 dB for $\Delta f = $ 60 kHz and $\Delta f = $ 120 kHz, respectively. Besides, compared to the RCRB, the proposed algorithm suffers from only 1.5 dB and 1.3 dB SNR degradation for $\Delta f = $ 60 kHz and $\Delta f = $ 120 kHz, respectively. For the 3D-DFT method, since the performance of azimuth estimation is limited by the array aperture of receive antennas, the azimuth RMSE does not change when SNR $>$ -15 dB. Moreover, when the subcarrier spacing is  $\Delta f =$ 60 kHz, the proposed algorithm achieves almost the same azimuth estimation performance as that for $\Delta f =$ 120 kHz.

\subsection{Estimation Accuracy for Multiple Targets}
\begin{figure}[!t]
\centering
\subfloat[RMSE of range estimation vs SNR.]{
\includegraphics[width=2.7in]{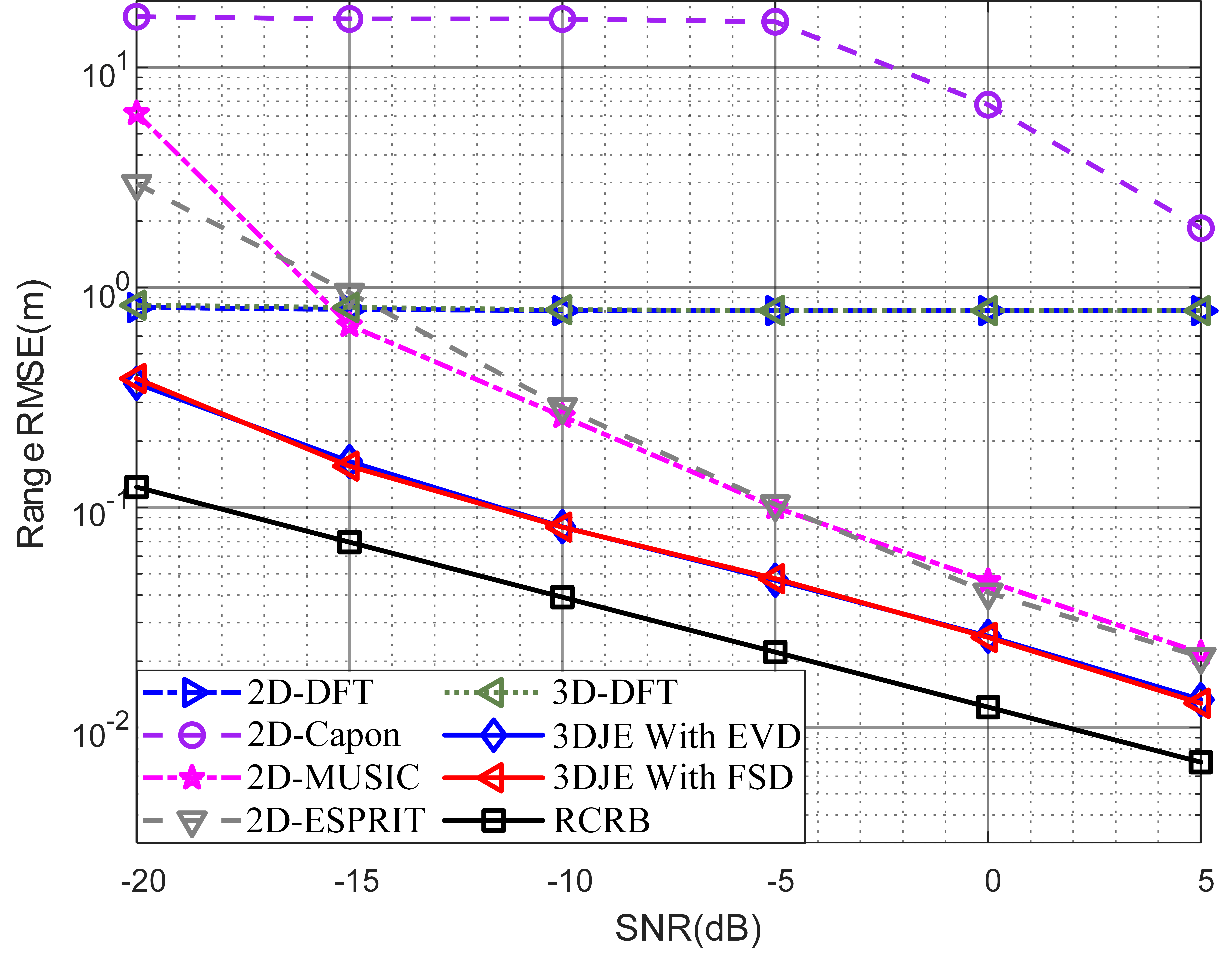} 
}

\subfloat[RMSE of velocity estimation vs SNR.]{
\includegraphics[width=2.7in]{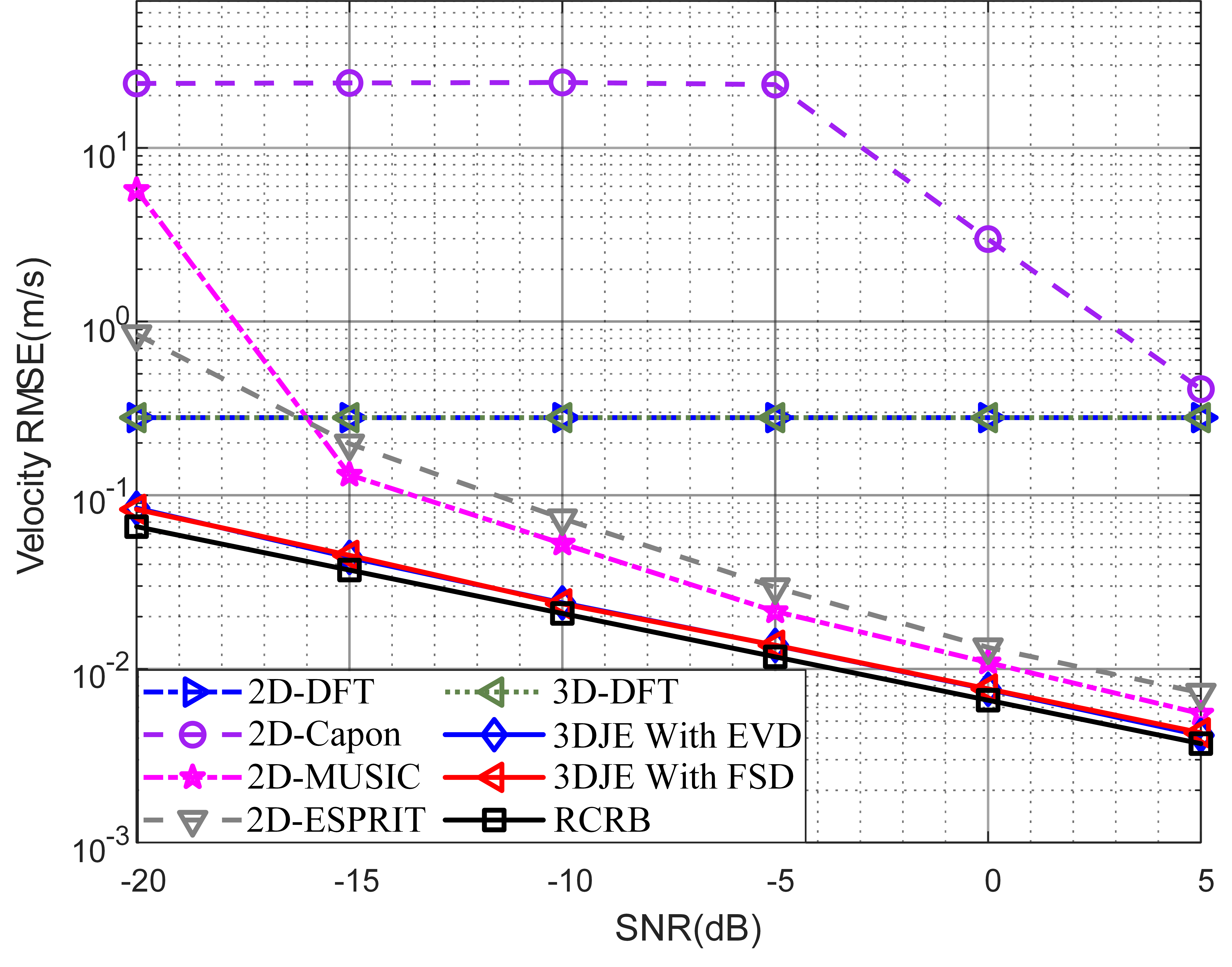} 
}

\subfloat[RMSE of azimuth estimation vs SNR.]{
\includegraphics[width=2.7in]{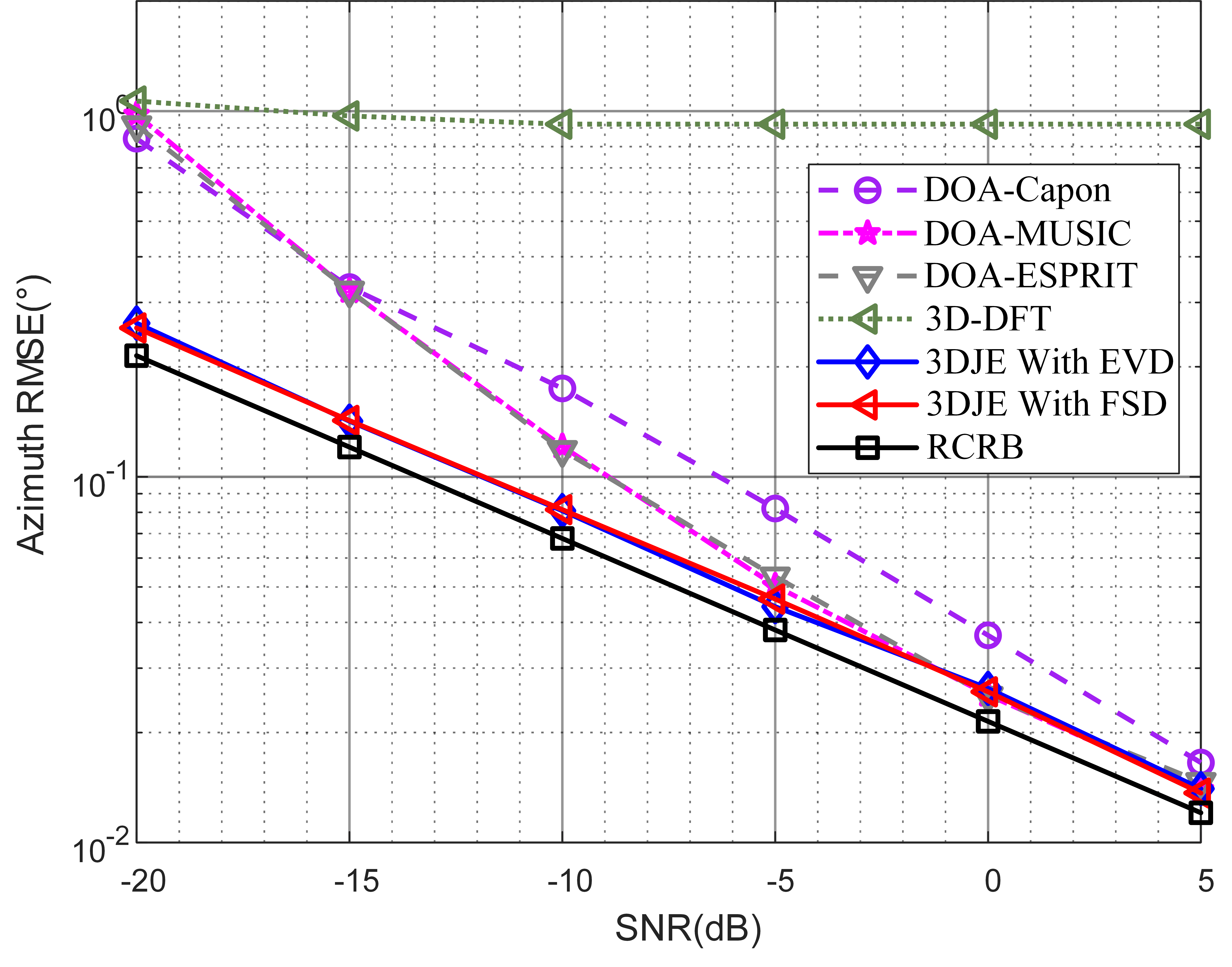} 
}

\caption{Multi-target estimation RMSE vs SNR for $\Delta f = 60\ {\rm{ }}\rm kHz$.}
\label{fig5}
\end{figure}

\begin{figure}[!t]
\centering
\subfloat[RMSE of range estimation vs SNR.]{
\includegraphics[width=2.7in]{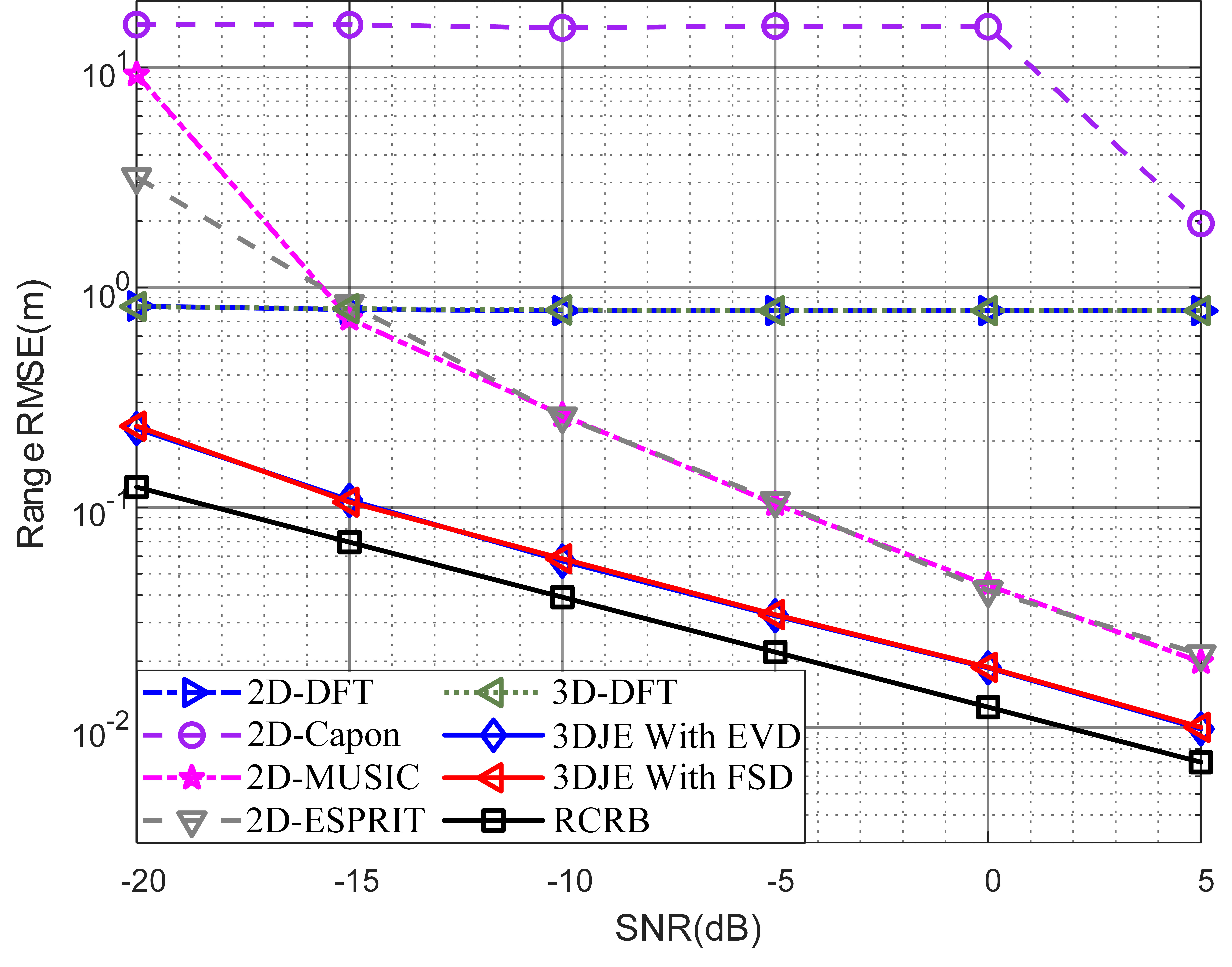} 
}

\subfloat[RMSE of velocity estimation vs SNR.]{
\includegraphics[width=2.7in]{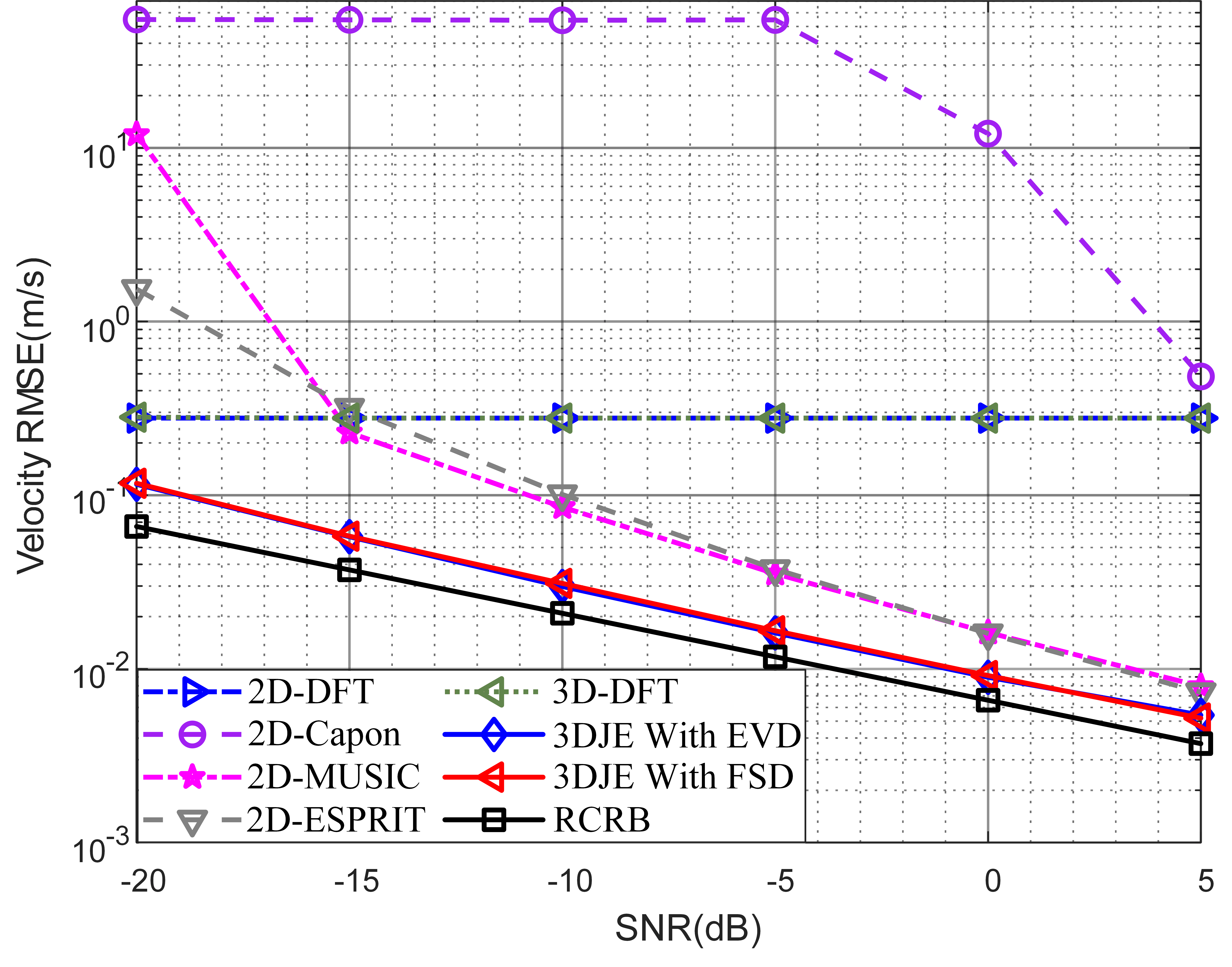} 
}

\subfloat[RMSE of azimuth estimation vs SNR.]{
\includegraphics[width=2.7in]{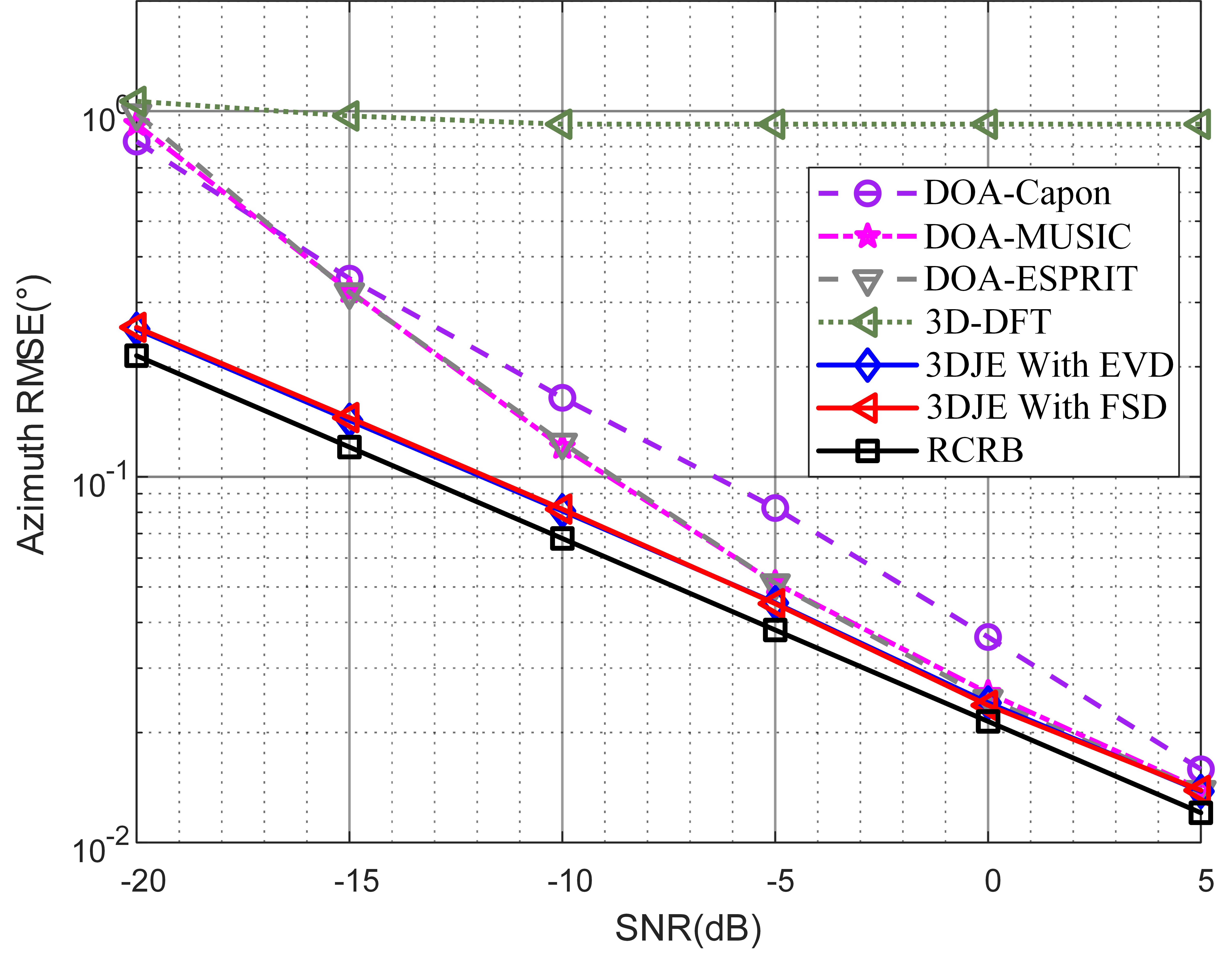} 
}

\caption{Multi-target estimation RMSE vs SNR for $\Delta f = 120\ {\rm{ }}\rm kHz$.}
\label{fig6}
\end{figure}

We set the number of targets $K=3$. The range, velocity, and azimuth parameters of the three targets are assumed to be $(35{\rm{m, 15m/s, 20}}^\circ )$, $(60{\rm{m, 10m/s, -20}}^\circ )$ and $(80{\rm{m, -10m/s, 50}}^\circ )$, respectively. The SNR varies from -20 dB to 5 dB with an interval of 5 dB. For each SNR, 200 times independent Monte Carlo simulations are performed. The simulation results are shown in Fig. \ref{fig5} and Fig. \ref{fig6}. Numerical results show that the proposed 3DJE algorithm with FSD achieves almost the same RMSE performance as the 3DJE with EVD.

In Fig. \ref{fig5}(a) and Fig. \ref{fig6}(a), the RMSE of range estimation with different SNRs is illustrated. The proposed algorithm achieves better estimation performance than other algorithms, especially when the SNR is low. Specifically, at a RMSE level of 10$^{-1}$, compared to the 2D-MUSIC method, the proposed algorithm achieves a SNR gain of 6.6 dB and 9.6 dB for $\Delta f = $ 60 kHz and $\Delta f = $ 120 kHz, respectively. Besides, for $\Delta f = $ 120 kHz, the proposed algorithm suffers from about 3.6 dB SNR degradation compared to the RCRB. Moreover,  the proposed algorithm achieves better range estimation performance for $\Delta f =$ 120 kHz than that for $\Delta f =$ 60 kHz.

In Fig. \ref{fig5}(b) and Fig. \ref{fig6}(b), the RMSE of velocity estimation with different SNRs is presented. Fig. \ref{fig5}(b) and Fig. \ref{fig6}(b) show that the proposed 3DJE algorithm achieves better estimation performance than other algorithms, especially when the SNR is low. Specifically, at a RMSE level of 10$^{-2}$, compare to the RCRB, the proposed algorithm suffers from only 1.3 dB and 2.8 dB SNR degradation for $\Delta f = $ 60 kHz and $\Delta f = $ 120 kHz, respectively. Moreover, the proposed algorithm achieves better velocity estimation performance for $\Delta f = $ 60 kHz than that for $\Delta f =$ 120 kHz.

In Fig. \ref{fig5}(c) and Fig. \ref{fig6}(c), the RMSE of azimuth estimation with different SNRs is demonstrated. Fig. \ref{fig5}(c) and Fig. \ref{fig6}(c) show that the proposed 3DJE algorithm achieves better estimation performance than other algorithms, especially when the SNR is low. Specifically, at a RMSE level of 10$^{-1}$, compared to the 2D-MUSIC method, the proposed algorithm achieves a SNR gain of 3 dB and 2.9 dB for $\Delta f = $ 60 kHz and $\Delta f = $ 120 kHz, respectively. Besides, compared to the RCRB, the proposed algorithm suffers from only 1.5 dB and 1.6 dB SNR degradation for $\Delta f = $ 60 kHz and $\Delta f = $ 120 kHz, respectively. Moreover, when the subcarrier spacing is  $\Delta f = $ 60 kHz, the proposed algorithm achieves almost the same azimuth estimation performance as that for $\Delta f =$ 120 kHz.

\begin{figure}[!ht]
\centering
\includegraphics[width=2.7in]{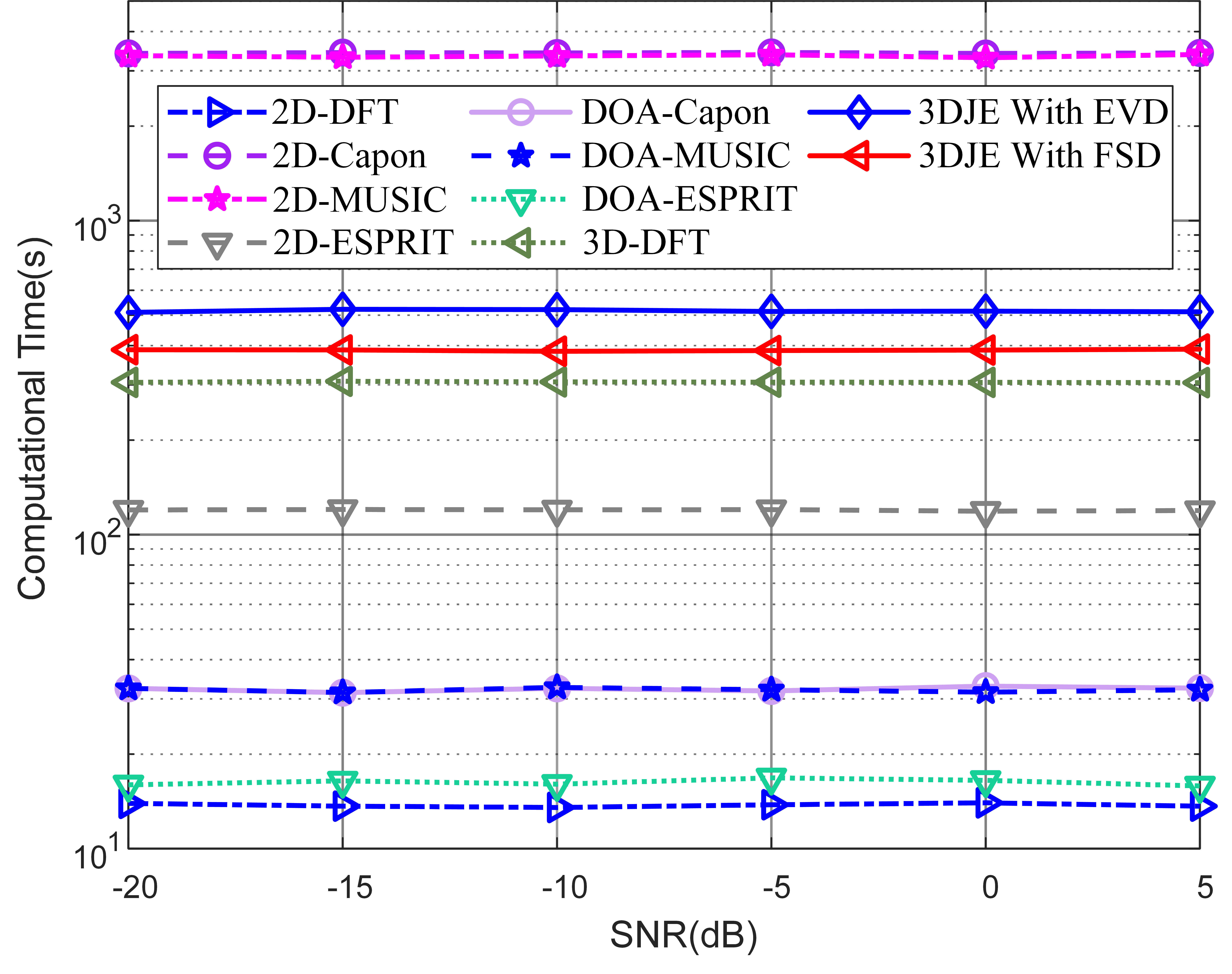}
\caption{Computational time vs SNR.}
\label{fig7}
\end{figure}

Fig. \ref{fig7} shows the computational time of 200 Monte Carlo simulations with different SNRs. With the increase of SNR, the computational time of the algorithms is almost unchanged. The computational time of the 3DJE with EVD and 3DJE with FSD is about 515.6 s and 387.0 s, respectively. Compared with 3DJE with EVD, 3DJE with FSD reduces approximately 25.0\% of computational time and has almost no performance loss. Besides, compared with the 2D-DFT method, the 3D-DFT method requires additional DFT operation in the space domain, which leads to a sharp increase in computational time. In addition, since the 2D-MUSIC and 2D-Capon methods need to search the spectral peaks, they have a longer computational time than the 2D-ESPRIT and 2D-DFT methods. Similarly, the DOA-Capon and DOA-MUSIC methods also have longer computational time than the DOA-ESPRIT method.

\section{Conclusion}
This paper proposes an auto-paired super-resolution 3DJE algorithm to achieve the joint range-velocity-azimuth estimation for OFDM-based ISAC systems. Simulation results show that compared to the classic 2D-DFT, 2D-MUSIC, 2D-Capon, 2D-ESPRIT and 3D-DFT methods, the proposed algorithm achieves better range and velocity estimation performance. Compared to the classic DOA estimation methods, including DOA-ESPRIT, DOA-MUSIC,  DOA-Capon and 3D-DFT, the proposed algorithm achieves better azimuth estimation performance. Besides,  the proposed algorithm suffers slightly RMSE performance degradation compared to the RCRB. For future work, we will consider more complicated situations with non-Gaussian white noise, and elevation estimation for OFDM-based ISAC systems.

{\appendix[Proof of Lemma 1]}
Let $\mathbf w'_{p,m}=[ {w'_{p,m}(0)}, {w'_{p,m}(1)},\ldots, {w'_{p,m}(N\!-\!1)}]^T$ and $\mathbf w_{p,m}=[w_{p,m}(0),w_{p,m}(1),\ldots,w_{p,m}(N\!-\!1)]^T$. Assume that the joint probability density distribution function (PDF) of $\mathbf w'_{p,m}$ and $\mathbf w_{p,m}$ are $ f(\mathbf w')$ and $ f(\mathbf w)$, respectively. $\mathbf F$ is defined as the Fourier transform matrix, which can be expressed as
\begin{equation}
\label{deqn_ex75}
\mathbf F = \left[ {\begin{array}{*{20}{c}}
1&1&1& \cdots &1\\
1&\zeta &{{\zeta ^2}}& \cdots &{{\zeta ^{N - 1}}}\\
1&{{\zeta ^2}}&{{\zeta ^4}}& \cdots &{{\zeta ^{N - 2}}}\\
 \vdots & \vdots & \vdots & \ddots & \vdots \\
1&{{\zeta ^{N - 1}}}&{{\zeta ^{N - 2}}}& \cdots &\zeta 
\end{array}} \right],
\end{equation}
where $\zeta = {{e^{ - j{{2\pi } \mathord{\left/{\vphantom {{2\pi } N}} \right.\kern-\nulldelimiterspace} N}}}}$.

Since ${w'_{p,m}}({k})$ is independent and ${w'_{p,m}}({k})\sim {\cal C}{\cal N}(0 ,{\sigma ^2})$, it is obvious that $\mathbf w'_{p,m} \sim {\cal C}{\cal N}(\mathbf 0, {\sigma ^2}\mathbf I)$. The joint PDF $ f(\mathbf w')$ can be expressed as  
\begin{equation}
\label{deqn_ex76}
f({\mathbf{w'}}) = \frac{1}{{{{(\pi {\sigma ^2})}^N}}}\exp \left\{- \displaystyle \frac{{\left\| {{\mathbf{w'}}} \right\|_2^2}}{{{\sigma ^2}}}\right\}.
\end{equation}

Since $\mathbf w_{p,m} = \mathbf F \mathbf w'_{p,m}$, we have $\mathbf w_{p,m} \sim {\cal C}{\cal N}(\mathbf 0 ,{\sigma ^2}\mathbf F \mathbf F^H)$, and the joint PDF $ f_{w}(\mathbf w)$ can be expressed as 
\begin{equation}
\label{deqn_ex77}
{f}(\mathbf w) = \frac{1}{{{\pi ^N}\det ({\sigma ^2}\mathbf F{\mathbf F^H})}}\exp \left\{ - \frac{{{\mathbf w^H}(\mathbf F{\mathbf F^H)^{-\!1}}\mathbf w}}{{{\sigma ^2}}}\right\}.
\end{equation}

Moreover, since the $\mathbf F \mathbf F^H = N \mathbf I$, (\ref{deqn_ex77}) can be  further written as
\begin{equation}
\label{deqn_ex78}
{f}(\mathbf w) = \frac{1}{{{{(\pi N{\sigma ^2})}^N}}}\exp \left\{ - \frac{{{{\left\| \mathbf w \right\|}_2^2}}}{{N{\sigma ^2}}}\right\}.
\end{equation}

The PDF in (\ref{deqn_ex78}) shows that the random variable ${w_{p,m}}({n})$ obeys a complex Gaussian distribution, denoted by ${w_{p,m}}({n}) \sim {\cal C}{\cal N}( 0,N{\sigma ^2})$. In addition, the random variables ${w_{p,m}}({n})$ with different frequencies are independent.

Suppose that the modulation order of PSK is $M_{{p}}$, \vspace{0.05cm} the PSK modulation symbol $s_m(n)$ can be expressed as $s_m(n)={e^{j\frac{{2\pi q}}{{{M_{p}}}}}}=\cos (\frac{{2\pi q}}{{{M_{p}}}}) + j\sin (\frac{{2\pi q}}{{{M_{p}}}})=s_{re}+js_{im}=s$,\vspace{0.05cm}  where $q=0,1,\ldots,M_{{p}}\!-\!1$.  It is obvious that $\sum\limits_{q = 0}^{{M_p\! - \!1}} {P({s_{re}} = \cos \frac{{2\pi q}}{{{M_{p}}}},{s_{{im}}} = \sin \frac{{2\pi q}}{{{M_{p}}}})}=1$.

For convenience, let ${w_{p,m}}(n)=w_{re}+jw_{im}=w $ and ${v_{p,m}}(n)=v_{re}+jv_{im}=v$, since $v=w/s$, we have
\begin{equation}
\label{deqn_ex79}
\begin{array}{ll}
v &=\displaystyle \frac{w}{s} = \displaystyle \frac{{{w_{re}} + j{w_{im}}}}{{{s_{re}} + j{s_{im}}}} \vspace{0.05cm}\\
 &= ({w_{re}}{s_{re}} + {w_{im}}{s_{im}}) + j({w_{im}}{s_{re}} - {w_{re}}{s_{im}})\\
 &= {v_{re}} + j{v_{im}}.
\end{array}
\end{equation}

The PDF of $v_{re}$ can be expressed as 
\begin{equation}
\label{deqn_ex80}
\begin{array}{ll}
f({v_{re}}) = \sum\limits_{q = 0}^{{M_{p} \!-\! 1}} {P({s_{re}} = \cos \frac{{2\pi q}}{{{M_{p}}}},{s_{im}} = \sin \frac{{2\pi q}}{{{M_{p}}}})} \\ 
 \cdot \displaystyle \frac{1}{{\sqrt {\pi (s_{re}^2 + s_{im}^2)N{\sigma ^2}{\text{ }}} }}\exp \left\{ { - \frac{{v_{re}^2}}{{(s_{re}^2 + s_{im}^2)N{\sigma ^2}{\text{ }}}}} \right\} \\
 \ \ \ \ \ \ \ = \displaystyle \frac{1}{\sqrt {\pi N{\sigma ^2}}} \exp \left\{ - \frac{v_{re}^2}{N{\sigma ^2}} \right\}.
\end{array}
\end{equation}

Similarly, we can obtain that ${v_{im}}\sim {\cal N}(0 ,N{\sigma ^2}/2)$. In addition, since ${\rm \mathbb{E}}\{ {v_{re}}{v_{im}}\}  = 0$, $v_{re}$ and $v_{im}$ are mutually independent. Therefore, ${v_{p,m}}(n)=v_{re}+jv_{im}$ obeys a complex Gaussian distribution, denoted by ${v_{p,m}}(n)\sim {\cal C}{\cal N}(0 ,N{\sigma ^2})$.

\end{document}